\documentclass[journal]{IEEEtran}
\usepackage{graphicx}
\usepackage{comment}
\usepackage{amsmath,amssymb} 
\usepackage{color}
\usepackage{array}
\usepackage{multirow}
\usepackage{subfigure}
\usepackage{balance}\usepackage{hyperref}
\hypersetup{    
    urlcolor=magenta,
}
\ifCLASSINFOpdf
\else
\fi

\begin{document}
\title{Hyperspectral Image Super-resolution via Deep
	Progressive Zero-centric Residual Learning}

\author{Zhiyu Zhu, Junhui Hou, \textit{Senior Member, IEEE}, Jie Chen, \textit{Member, IEEE}, \\ Huanqiang Zeng, \textit{Senior Member, IEEE}, and Jiantao Zhou \textit{Senior Member, IEEE}
\thanks{This work was supported in part by the Hong Kong Research Grants Council under Grants 9048123 (CityU 21211518) and 9042820 (CityU 11219019), and in part by the Macau Science and Technology Development Fund under 077/2018/A2. (\textit{Corresponding Author: Junhui Hou})}
\thanks{Z. Zhu and J. Hou are with the Department of Computer Science, City University of Hong Kong, Hong Kong. E-mail: zhiyuzhu2-c@my.cityu.edu.hk and jh.hou@cityu.edu.hk}
\thanks{J. Chen is with the Department of Computer Science, Hong Kong Baptist University, Hong Kong. E-mail: chenjie@comp.hkbu.edu.hk.}
\thanks{H. Zeng is with the School of Information Science and Engineering, Huaqiao University, Xiamen 361021, China. E-mail: zeng0043@hqu.edu.cn}
\thanks{J. Zhou is with the Department of Computer and Information Science,  University of Macau, Macau. E-mail: jtzhou@um.edu.mo}}

\markboth{IEEE Transactions on Image Processing}%
{Shell \MakeLowercase{\textit{et al.}}: Bare Demo of IEEEtran.cls for IEEE Journals}
\maketitle
\begin{abstract}
This paper explores the problem of hyperspectral image (HSI) super-resolution that merges a low resolution HSI (LR-HSI) and a high resolution multispectral image (HR-MSI). The cross-modality distribution of the spatial and spectral information makes the problem challenging. Inspired by the classic wavelet decomposition-based image fusion, we propose a novel \textit{lightweight} deep neural network-based framework, namely progressive zero-centric residual network (PZRes-Net), to address this problem efficiently and effectively. Specifically, PZRes-Net learns a high resolution and \textit{zero-centric} residual image, which contains high-frequency spatial details of the scene across all spectral bands, from both inputs in a progressive fashion along the spectral dimension. And the resulting residual image is then superimposed onto the up-sampled LR-HSI in a \textit{mean-value invariant} manner, leading to a coarse HR-HSI, which is further refined by exploring the coherence across all spectral bands simultaneously. To learn the residual image efficiently and effectively, we employ spectral-spatial separable convolution with dense connections. In addition, we propose zero-mean normalization implemented on the feature maps of each layer to realize the zero-mean characteristic of the residual image. Extensive experiments over both real and synthetic benchmark datasets demonstrate that our PZRes-Net outperforms state-of-the-art methods to a \textit{significant} extent in terms of both 4 quantitative metrics and visual quality, e.g., our PZRes-Net improves the PSNR more than 3dB, while saving 2.3$\times$ parameters and consuming 15$\times$ less FLOPs. The code is publicly available at \url{https://github.com/zbzhzhy/PZRes-Net}
 
\end{abstract}


\IEEEpeerreviewmaketitle

\section{Introduction}
\IEEEPARstart{H}{yperspectral} imaging is aimed at collecting information from across the electromagnetic spectrum for each pixel of the image of a scene \cite{bioucas2013hyperspectral,khan2018modern}. The rich spectral information of the recorded hyperspectral image (HSI) enables it to deliver more faithful knowledge of a targeted scene than conventional imaging modalities \cite{yan2019cross}. As a result, the HSI has grown increasingly popular over the past ten years in various fields, such as military, industrial, and scientific arenas. 
The HSI has also boosted the performance of applications in computer vision, e.g.,
binary partition tree-based HSI segmentation \cite{veganzones2014hyperspectral}, graph convolutional neural network-based HSI classification \cite{hong2020graph}, an augmented linear minxing model for hyperspectral unmixing \cite{hong2018augmented}, hierarchical analysis-based object tracking in HSIs \cite{tochon2017object}, semi-supervised HSI segmentation via the Bayesian approach and multinomial logistic regression \cite{li2010semisupervised,li2011hyperspectral}, etc.

\begin{figure}
     \centering
     \includegraphics[width=0.45\textwidth]{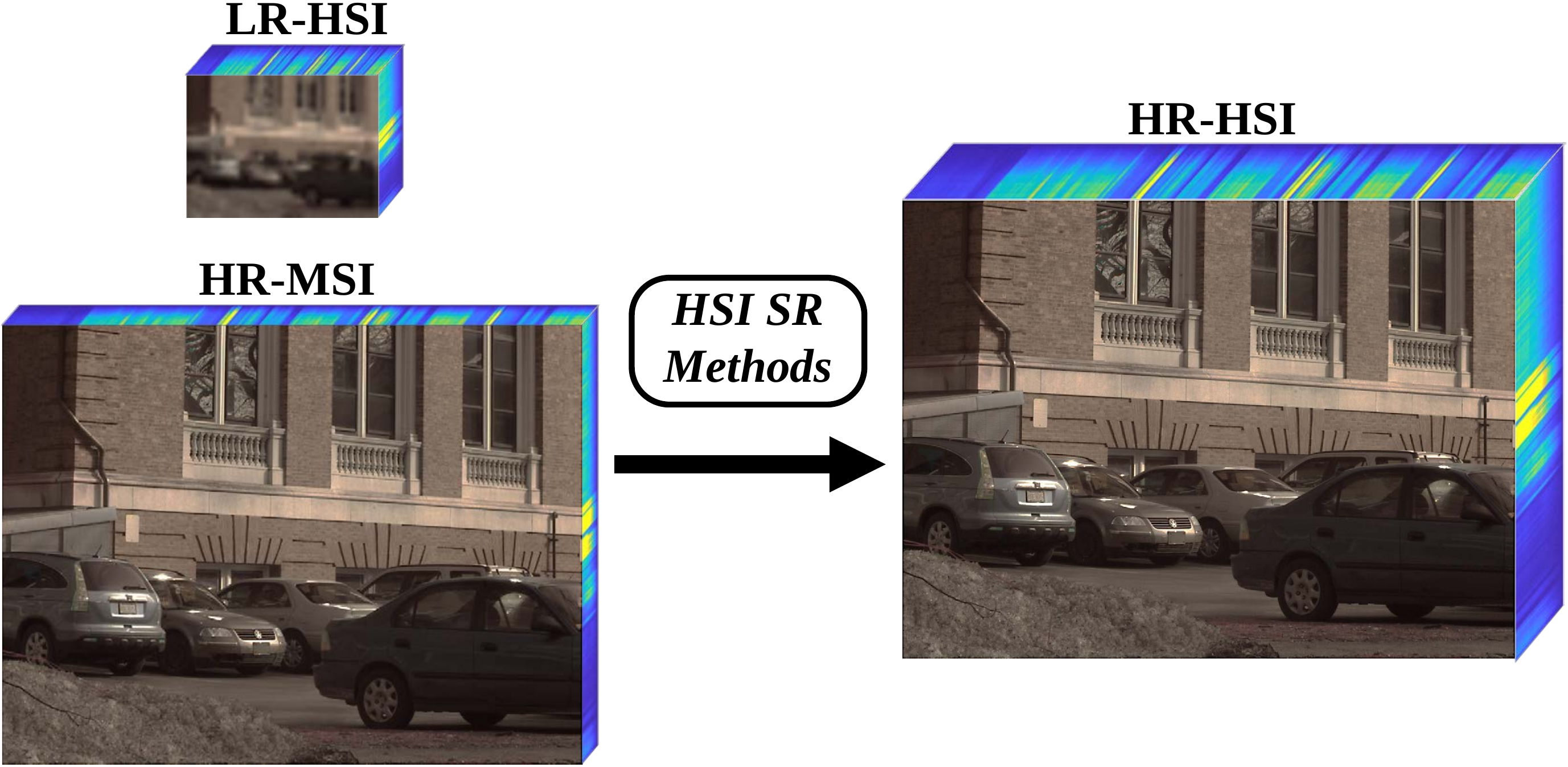}
     \caption{The illustration of HSI super-resolution from an LR-HSI and an HR-MSI that capture the same scene.}
     \label{fig:fusion}
 \end{figure}

However, due to the hardware limitation of existing imaging systems, there is an inevitable trade-off between the spectral and spatial resolution \cite{dian2018deep}. For a specific optical system, it could only record the image with either high spatial resolution together with very limited spectral bands, e.g., the high resolution multispectral image (HR-MSI), or dense spectral bands with reduced spatial resolution, e.g., the low resolution HSI (LR-HSI). Hence, as illustrated in Fig.\ref{fig:fusion}, HSI super-resolution (a.k.a MSI/HSI fusion) that merges an HR-MSI and an LR-HSI has being become a promising way to obtain  HR-HSIs \cite{xie2019multispectral,wang2019deep,dian2018deep}.  

To tackle this challenge, various methods have been proposed in the last few decades. From the perspective of signal processing, multi-scale decomposition-based methods that only consume very limited computation resources have demonstrated their abilities in information fusion \cite{li2017pixel}, e.g., pyramid-based \cite{kautz2007exposure} and wavelet decomposition-based \cite{nunez1999multiresolution,zhang1999categorization}.
Based on the prior knowledge of HSIs, e.g., the sparse prior and the low-rank prior, plenty of matrix factorization based-methods have emerged  \cite{zhang2016multispectral,dong2016hyperspectral,grohnfeldt2013jointly}. Recently, owing to the remarkable representation learning ability, deep neural network (DNN)-based methods have been introduced 
\cite{xie2019multispectral,dian2018deep,wang2019deep,fu2018hyperspectral}, \cite{zheng2020hyperspectral}, whose performance exceeds that of traditional/non-DNN methods to a large extent
(see Section \ref{sec:RW} for more details). However, the reconstruction quality of the current state-of-the-art methods is still not satisfactory, due to the insufficient utilization/modeling of the cross-modality information.
 
Inspired by the classic wavelet decomposition-based methods \cite{nunez1999multiresolution,gonzalo2004customized}, we propose a novel DNN-based framework, namely \textit{Progressive Zero-centric Residual Network} (PZRes-Net), to achieve HSI super-resolution in both \textit{efficient} and  \textit{effective} ways. As shown in Fig. \ref{fig:CRseNet},  the input LR-HSI is first up-sampled in a \textit{mean-value invariant} manner. Following that, a \textit{zero-centric} residual image is progressively learned along the spectral dimension from both the up-sampled LR-HSI and HR-MSI with a zero-centric residual learning module, in which spectral-spatial separable convolutions with dense aggregation extract spectral-spatial information efficiently and effectively. Moreover, zero-mean normalization is applied for promoting the zero-centric characteristic of the learned residual image.  The resulting residual image is further superimposed on the up-sampled LR-HSI, leading to a coarse HR-HSI, which is finally refined through exploring the coherency among all spectral bands simultaneously. We conduct various and extensive experiments and comparisons to evaluate and analyze the proposed PZRes-Net comprehensively. It is concluded that our PZRes-Net remarkably outperforms state-of-the-art methods both quantitatively and qualitatively across multiple real and synthetic benchmark datasets. Especially, our PZRes-Net improves the PSNR more than 3dB, while saving 2.3$\times$ parameters and consuming 15$\times$ less FLOPs.

The rest of this paper is organized as follows. Section \ref{sec:RW} briefly reviews existing methods. Section \ref{sec:PM} presents the proposed method, namely PZRes-Net, followed by extensive experimental results as well as comprehensive analyses on both synthetic and real data in Section \ref{sec:EXP}. Finally Section \ref{sec:CON} concludes this paper. 

\section{Related Work}
\label{sec:RW}
We classify existing methods into two categories, i.e., 
(1) traditional methods, including multi-scale decomposition-based 
\cite{kautz2007exposure}, 
\cite{zhang1999categorization,pajares2004wavelet,li2002using}, 
\cite{yang2012fusion,zheng2007multisource} 
and optimization-based 
\cite{akhtar2014sparse}; and (2) deep learning-based methods \cite{dian2018deep,xie2019multispectral,wang2019deep}. In the following, we will review them in detail.

\subsection{Traditional Methods}
Multi-scale decomposition-methods focus on representing the image spatial structures with multiple layers. Various wavelet decomposition (WD)-based methods have been proposed in the past few decades \cite{nunez1999multiresolution,kautz2007exposure,zhang1999categorization,nunez1997simultaneous,yocky1996multiresolution,nunez1999image}. For example, Nunez \textit{et al.} \cite{nunez1999multiresolution} proposed two types of WD-based methods for MSI pansharpening: the additive method and the substitution method. Specifically, the former consists of the following steps: (1) register an LR-MSI with an HR-panchromatic image and upsmaple it to the same resolution in order to be superimposed;
(2) decompose the HR-panchromatic image to several wavelet planes containing high-frequency spatial information which follows zero-mean distribution; and (3) add the wavelet planes to the up-sampled LR-MSI. However, the latter decomposes both inputs, then replace wavelet planes of the up-sampled LR-MSI with those of the HR-panchromatic images. They also experimentally demonstrated that the additive method preserves spatial information better than the substitution method. Gonzalo \textit{et al.} \cite{gonzalo2004customized} also proposed an adaptive WD-based method, which fuses the wavelet planes 
with different weights.

Based on the prior knowledge of HSIs \cite{xie2019multispectral}, a considerable number of traditional machine learning-based methods have been proposed. 
For example,  matrix factorization-based methods assume that each spectrum can be linearly represented with several spectral atoms \cite{dian2019learning}. Under the assumption that an HSI lies in a low-dimensional subspace,  Wei \textit{et al.} \cite{wei2015hyperspectral} 
used the spatial dictionaries learned from the HR-MSI to promote the spatial similarities. Fang \textit{et al.} \cite{fang2018super} proposed a super-pixel-based sparse representation. Han \textit{et al.} \cite{han2018self} utilized a self-similarity prior as the constraint for the sparse representation of the input HSI and MSI. Akhtar \textit{et al.} \cite{akhtar2014sparse} first learned a non-negative dictionary, then introduced a simultaneous greedy pursuit algorithm to estimate coefficients for each local patch. Dong \textit{et al.} \cite{dong2016hyperspectral} proposed a matrix factorization-based method which imposes joint sparsity and non-negativity constraints on the learned representation coefficients. Through tensorization techniques, an image in the form of the conventional 2-D matrix can be converted to a 4-D or even higher-order tensor without loss of information. Thus, some tensor factorization-based methods have also been proposed to address the fusion problem. For example, Kanatsoulis \textit{et al.} \cite{kanatsoulis2018hyperspectral} established a coupled tensor factorization framework. Xu \textit{et al.} \cite{xu2020hyperspectral} utilized the high-order coupled tensor ring representation with graph-Laplacian regularization to realize HSI super-resolution.

However, these traditional methods were usually constructed based on some priors, e.g., sparse prior, low-rank prior, and global similarity prior,  which may not be consistent with the complex real world scenarios \cite{xie2019multispectral}.

\subsection{DNN-based Methods}
Powered by the strong representations learning ability, DNNs have become an emerging tool for HSI super-resolution \cite{dian2019learning}. Palsson \textit{et al.} \cite{palsson2017multispectral} proposed a 3-D convolutional neural network (CNN)-based MSI/HSI fusion approach and reduced the computational cost by using principal component analysis (PCA). Dian \textit{et al.} \cite{dian2018deep} used deep priors learned by residual learning-based DNNs and reconstructed HR-HSI by solving optimization problems. Mei \textit{et al.} \cite{mei2017hyperspectral} proposed a 3-D CNN to exploit both the spatial context and the spectral correlation. Arun \textit{et al.} \cite{arun2018cnn} explored DNNs to jointly optimize the unmixing and mapping operations in a supervised manner. Xie \textit{et al.} \cite{xie2019multispectral,xie2020mhf} proved that an HR-HSI could be represented by the linearly transformed HR-MSI and a to-be-estimated residual image, then unfolded an iterative algorithm, which solves aforementioned two components within a deep learning framework for HSI/MSI fusion. Zheng \textit{et al. \cite{zheng2020coupled} proposed an unsupervised coupled CNN with an adaptive response function for HSI super-resolution. Arun \textit{et al.} \cite{arun2020cnn} proposed a 3-D CNN-based HSI super-resolution, where novel hypercube-specific loss functions are used to augment the learning capability of the network.} Wang \textit{et al.} \cite{wang2018hyperreconnet,wang2015dual} tried to solve the HSI reconstruction using compressive sensing based methods. Xiong \textit{et al.} \cite{xiong2017hscnn} proposed CNN-based methods to achieve the recovery of HSIs from RGB images.  Qu \textit{et al.} \cite{qu2018unsupervised} solved the HSI super-resolution problem using an unsupervised encoder-decoder architecture.

Most of the above-mentioned methods are not blind, meaning that they are trained with the knowledge of HR-HSIs or degradation models. Although Wang \textit{et al.} \cite{wang2019deep} proposed a blind fusion model similar to \cite{xie2019multispectral}, their performance leaves much to be desired. Moreover, due to the iterative HSI refinement, their computation burdens are very high, which may restrict practical implementations.

\begin{figure*}
	\centering
	\includegraphics[width=18cm,height=9.5cm]{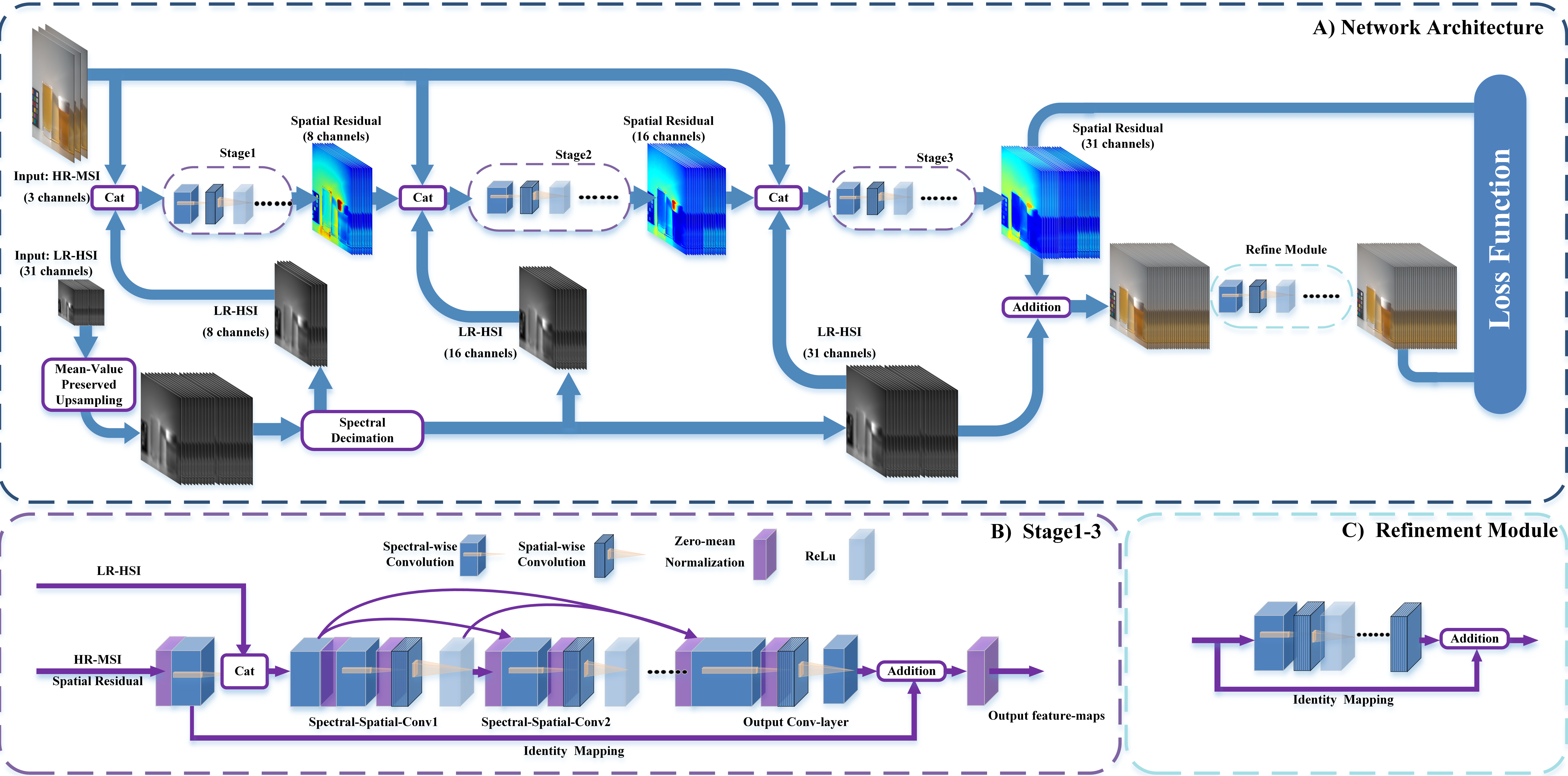}
	\caption{(a) The overall flowchart of the proposed PZRes-Net for HSI super-resolution that merges an LR-HSI and an HR-MSI. (b) The network  architecture of a stage contained in the zero-centric residual learning module. (c) The network architecture of the refinement module.}
	\label{fig:CRseNet}
\end{figure*}

\section{The Proposed Method}
\label{sec:PM}
Let $\mathbf{X}\in\mathbb{R}^{HW\times S}$ be an $S$-spectral bands HR-HSI to be reconstructed, each column of which is the vectorial representation of a spectral band of spatial dimensions $H\times W$. The degradation models for the observed HR-MSI with $s$ ($s \ll S$) spectral bands denoted by $\mathbf{Y}\in\mathbb{R}^{HW \times s}$ and LR-HSI of spatial dimensions $h\times w$ denoted by $\mathbf{Z}\in \mathbb{R}^{hw \times S}$ ($h \ll H$, $w \ll W$) from $\mathbf{X}$ can be formulated as:
\begin{equation}
\mathbf{Y} = \mathbf{XR}+\mathbf{N}_y,
\label{hrmsi}
\end{equation}
\begin{equation}
\quad \mathbf{Z} = \mathbf{DB}\mathbf{X}+\mathbf{N}_z,
\label{lrhsi}
\end{equation}
where $\mathbf{R} \in \mathbb{R}^{S \times s}$ is the camera spectral response function that integrates over the spectral dimension of the HSI to produce the MSI;
$\mathbf{B} \in \mathbb{R}^{HW \times HW}$ represents the blurring matrix applied on the HR-HSI before performing spatial decimation via the matrix $\mathbf{D}\in \mathbb{R}^{hw\times HW}$; $\mathbf{N}_y\in\mathbb{R}^{HW\times s}$ and $\mathbf{N}_z\in\mathbb{R}^{hw\times S}$ are the noises in $\mathbf{Y}$ and $\mathbf{Z}$, respectively. From Eqs. (\ref{hrmsi}) and (\ref{lrhsi}), it can be known that  $\mathbf{Y}$ contains high-resolution spatial context, while $\mathbf{Z}$ keeps dense spectral details.
Thus, the challenge of HSI super-resolution, i.e., reconstructing $\mathbf{X}$ from  $\mathbf{Z}$ under the assistance of $\mathbf{Y}$, boils down to ``\textit{how to leverage the spatial advantage of $\mathbf{Y}$ and propagate it across the densely sampled spectral bands of $\mathbf{Z}$ effectively}."

\subsection{Motivation and Overview}
Multi-scale decomposition-based methods have demonstrated their effectiveness in image fusion 
\cite{zhou2016perceptual}, \cite{zheng2009multi}, \cite{li2017pixel}. Particularly, the classic wavelet decomposition-based scheme for enhancing an LR-image with an HR-image from another modality contains the following procedures: the LR-image is first up-sampled to the same resolution as the HR-image in order to be superimposed; and wavelet planes containing \textit{zero-centric/mean high-frequency} details are then obtained by decomposing the HR-image with a shift-invariant wavelet transform, which are further superimposed onto the up-sampled LR-image. Such a scheme is able to utilize the spatial detail information from both images. Moreover, as the wavelet planes are designed to have zero mean-values,  the total flux of the enhanced LR-image will be preserved. Inspired by the principles of traditional multi-scale decomposition-based methods, we will study the HSI super-resolution by exploring the powerful representation capabilities of DNNs to learn such zero-centric high-frequency details adaptively. In addition to inheriting the advantage of the traditional methods, it is expected that the \textit{data-adaptive} characteristic of such a learning manner can boost the performance, compared with the \textit{pre-defined} and \textit{data-independent} decomposition process in traditional frameworks.

\begin{figure}[t]
\centering
    \subfigure[DIV2K dataset]{
	\includegraphics[width=0.45\linewidth]{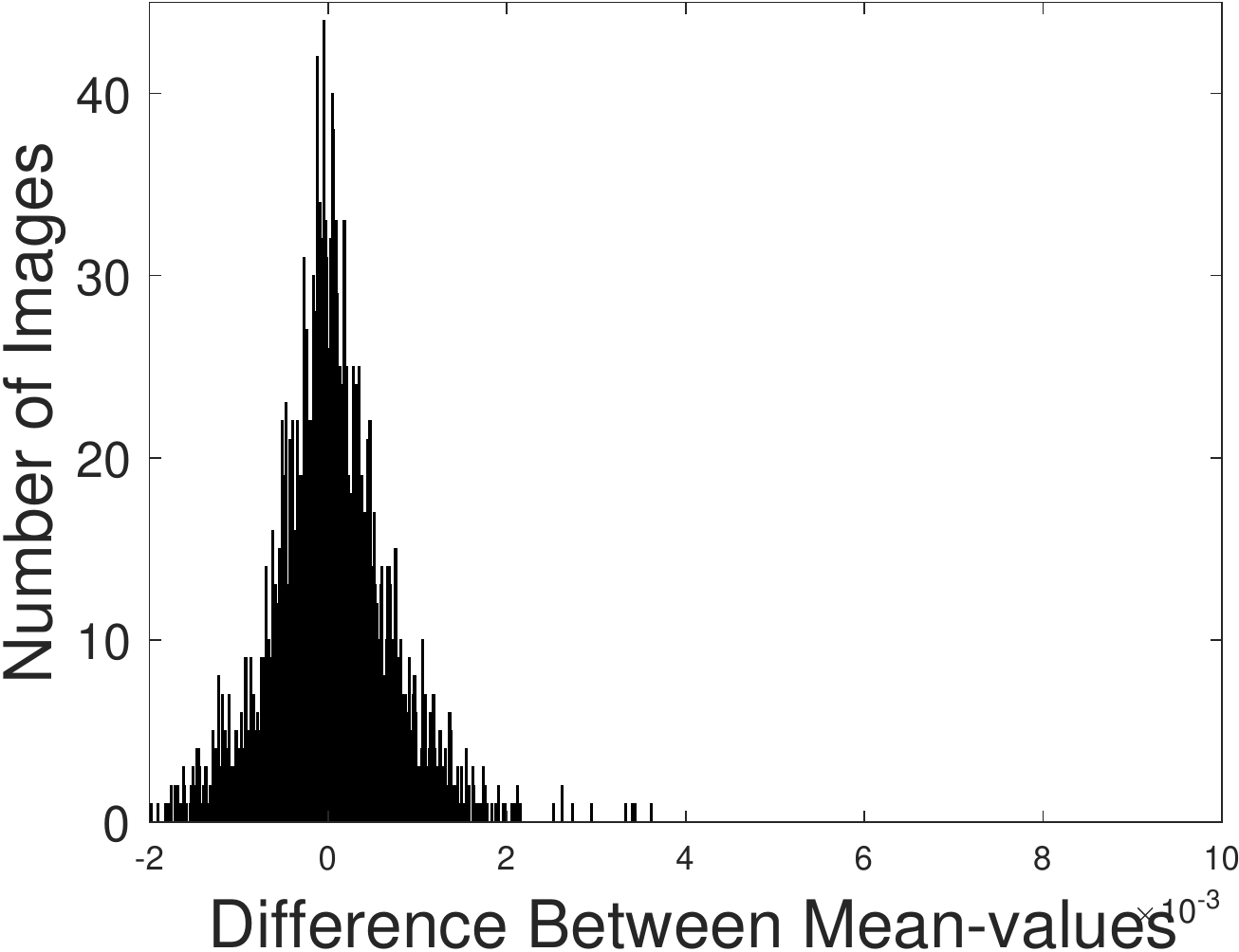} 
    }
    \subfigure[COCO dataset]{
	\includegraphics[width=0.45\linewidth]{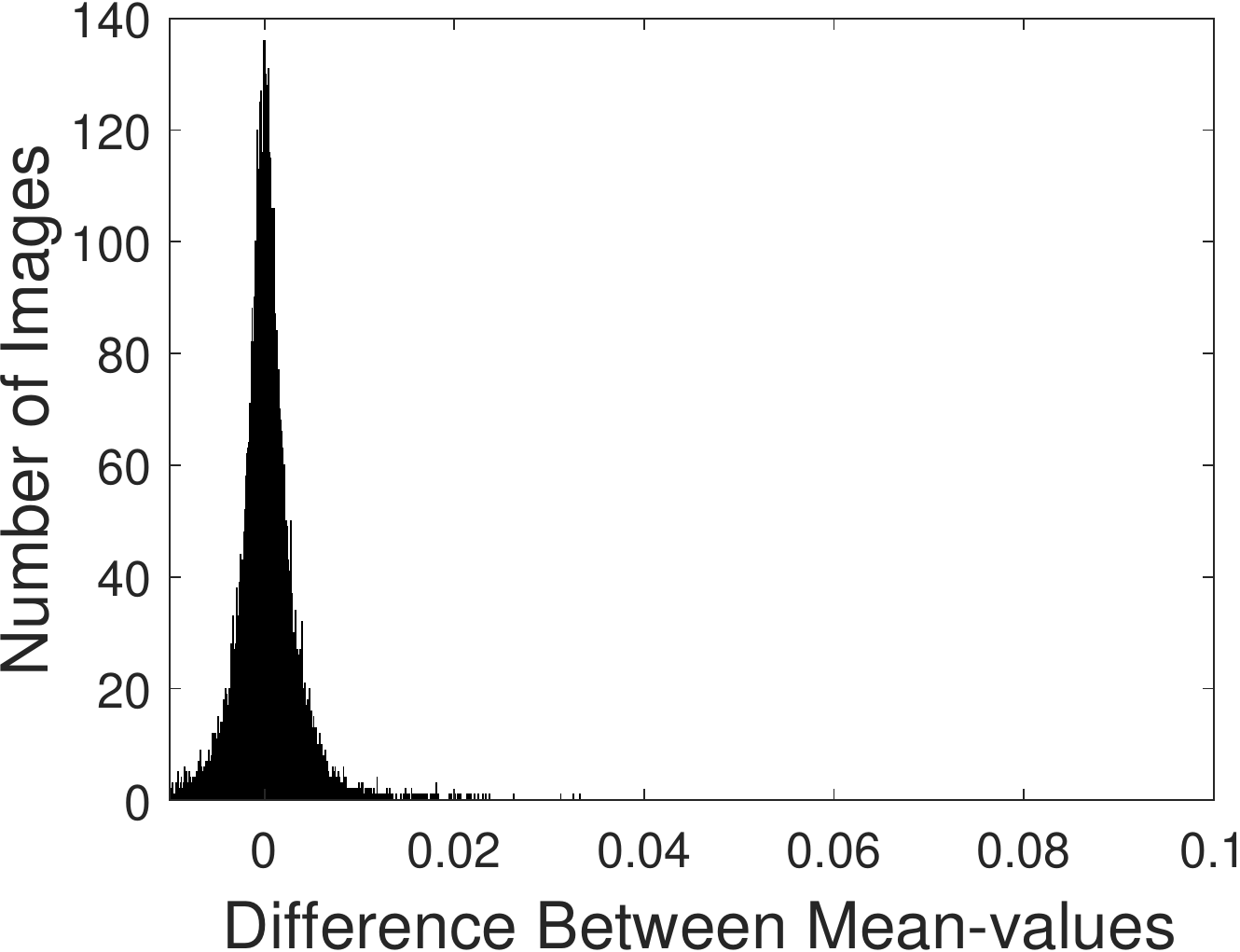} 
    }
\caption{Statistics of the difference between the mean-values of 
800 and 40K pairs of HR and LR RGB images respectively from DIV2K \cite{Agustsson_2017_CVPR_Workshops}, and COCO datasets \cite{lin2014microsoft}. All images are normalized to [0,1]. For each dataset, the LR-images were obtained by downsampling the corresponding HR ones with factor 8.
}
\label{fig:Spatial_prior}
\end{figure}

As shown in Fig. \ref{fig:CRseNet}, the proposed framework, namely progressive zero-centric residual network (PZRes-Net), is mainly composed of three modules: a mean-value invariant up-sampling module, a progressive zero-centric residual learning module, and a refinement module. Specifically, the up-sampling module first lifts the input LR-HSI to the same resolution as the input HR-MSI in a mean-value invariant manner. Then, the residual learning module estimates a residual image from both the input HR-MSI and up-sampled LR-HSI progressively along the spectral dimension in which \textit{zero-mean normalization} is applied on the input of each convolutional layer to enforce the mean-value of each band of the predicted residual image to be zero. The resulting \textit{zero-centric} residual image is further superimposed onto the up-sampled LR-HSI, leading to a coarse HR-HSI, which is finally fed into the refinement module, where the coherence across all spectral bands of the coarse HR-HSI is simultaneously explored in a residual learning manner for further augmenting reconstruction quality. In what follows, we will introduce each module in detail.

\textit{Remark}: It is worth pointing out the residual learning manner of our framework is essentially different from the well-known residual learning \cite{he2016deep} that is widely exploited in various networks and image/video applications \cite{zhang2020residual}, \cite{lan2020cascading}. 
The traditional residual learning was introduced with the purposes of facilitating neural network optimization or enhancing feature extraction, while such a manner of our PZRes-Net mimics the classic multi-scale decomposition-based fusion methods to realize information fusion. Moreover, our PZRes-Net learns a zero-centric residual, while there are no such constraints for traditional residual learning.

\subsection{Mean-value Invariant Up-sampling}
This module aims to lift the input LR-HSI into the same spatial resolution as the HR-MSI for subsequent residual superimposition. The \textbf{Law of Large Numbers} dictates that \textit{the observation average for a random variable should be close to its expectation value when it's based on a large number of trials}. Generally, image pixels from LR and HR captures are repetitive samples of the same real scene, and are expected to have an approximately identical mean-value which is aligned with the expectation value. Such an observation was also experimentally validated in Fig. \ref{fig:Spatial_prior}, where the histograms refer to the statistics of the difference between the mean values of 800 and 40K pairs of LR and HR RGB images respectively from DIV2K \cite{Agustsson_2017_CVPR_Workshops} and COCO \cite{lin2014microsoft}, two commonly-used benchmark image datasets. 

Recall that in our framework, a zero-mean residual image predicted by our zero-centric residual learning module will be superimposed on the up-sampled LR-HSI to form the HR-HSI. Therefore, to avoid distortion, the up-sampling process should be mean-value invariant, i.e., each band of the up-sampled LR-HSI will have an approximately identical mean-value to the corresponding band of the input LR-HSI. In order to achieve the objective, the widely-used transposed layer in image super-resolution could be employed but with additional restrictions on the learnable kernel installed in the layer, i.e., the sum of the elements of the kernel, which simultaneously convolve with the pixels of the input LR-image, should be equal to 1 while the kernel slides over the interpolated LR-image. For simplicity, in this paper we adopt the bi-linear interpolation to realize the mean-value invariant up-sampling, which has experimentally demonstrated to work very well. 

\subsection{Progressive Zero-centric Residual Learning}
In this module, a zero-centric residual image containing high-frequency spatial details of the captured scene is regressed by deeply extracting spatial context information from both the input HR-MSI and up-sampled LR-HSI. Inspired by the success of the progressive reconstruction strategy in image super-resolution \cite{ahn2018image, lai2017deep, karras2018progressive}, we embed the spectral bands of the up-sampled LR-HSI in a progressive fashion, rather than feed all of them into the network at the beginning. That is, as shown in Fig. \ref{fig:CRseNet}, the spectral bands of the up-sampled LR-HSI that are embedded into the network at different stages are regularly decimated with strides that decrease exponentially with the number of stages increasing. Taking an HSI with 31 spectral bands as an example, the numbers of decimated spectral bands from the first to the third stage are 8, 16, and 31, respectively.

Specifically, at each stage, 1-D convolution is first applied on the HR-MSI over the spectral domain to lift the number of feature channels to the same level as the input spectral bands decimated from the up-sampled LR-HSI, during which high-order details will be scattered to all channels. Then, the resulting feature maps from the HR-MSI and the decimated spectral bands from the up-sampled LR-HSI are concatenated along the spectral dimension, which are further fed into a sequence of spectral-spatial separable convolutional layers aggregated with densely connections \cite{huang2017densely,zhang2020residual} for efficient and comprehensive spectral-spatial feature extraction. During the feature extraction, the spatial details are mainly provided by the HR-MSI and propagated to all spectral bands with reference to the spectral information by the LR-HSI. Moreover, to obtain a zero-centric residual image, zero-mean normalization is applied on the input feature maps of each convolutional layer over the spatial dimension independently. In addition, an identity mapping which directly connects the output of the first spectral convolutional layer to the output of the stage in an additive manner to further enhance the flow of the high-frequency spatial details.  
Note that for stage-$i$ ($i>1$), the output of stage-$(i-1)$ is also concatenated as its input. 

Table \ref{tab:PZResNet_architecture} outlines the architecture details of the first stage. In the following, we will give more detailed elaborations towards the spectral-spatial separable convolution and zero-mean normalization.

\subsubsection{Spectral-spatial separable convolution with dense connections}

As mentioned earlier, the input to the residual learning module is a 3-D tensor, i.e., the concatenation of the 3-D HR-MSI and up-sampled LR-HSI. In order to comprehensively explore the information from both spectral and spatial domains, 3-D convolution is an intuitive choice to construct the residual learning module, which has demonstrated its effectiveness \cite{mei2017hyperspectral,li2017spectral}. However, compared with conventional 2-D convolution, 3-D convolution results in a significant increase in the parameter size, which may potentially cause over-fitting and consumption of huge computation resources. Analogy to the approximation of a high-dimensional filter with multiple low-dimensional filters in the field of signal processing, we use spectral-spatial separable (3S) convolutions to process the 3-D tensor for efficient spectral-spatial feature extraction. Note that separable convolutions have also demonstrated its effectiveness and efficiency in other deep learning-based image processing \cite{yeung2018light,wing2018fast,niklaus2017video}.

Specifically, the 3S convolution conducts 1-D spectral convolution (i.e., 1-D convolution over the spectral domain) and 2-D spatial convolution (i.e., independent 2-D convolution over the spatial domain of each feature map) sequentially. Also, there is an activation layer inserted between the two kinds of convolutions. The spectral convolution is equipped with a kernel of size $1\times C$ with $C$ being the number of feature channels, while the spatial convolution with a group of 2-D kernels of size $3\times 3$. Moreover, to enhance the feature extraction ability of the network for residual learning, we densely connect the 3S convolutional layers within a stage  \cite{huang2017densely}. That is, the feature maps obtained from all the preceding layers are concatenated along the spectral dimension and passed to the current layer. Additionally, such dense connections could potentially improve the information flow and reduce overfitting \cite{huang2017densely}.

\begin{table*}
	\caption{The architecture details of the first stage of the residual learning module and the refinement module. ``3S Convolution $j$" indicates the $j$-th spectral-spatial convolutional layer in the stage.}
	\centering
		\begin{tabular}{l c c c c c c}
			\hline	\hline
			$1^{st} stage$		&Kernel shape 	  						& \# Input Channels 	&\# Output Channels &Output shape               & ReLU & ZM-norm \\
			\hline
			\textbf{LinearTransform}	&8$\times$3$\times$1$\times$1			&3  					&8			 		&128$\times$128$\times$8	&--  		 &--	\\
			Concatenate		&--										&8+8=16  				&16			 		&128$\times$128$\times$16	&--  		 &--	\\
			\textbf{3S Convolution-$j$ $j\in [1,7]$	}\\
			Spectral Convolution										&16$j$ $\times$16$\times$1$\times$1			&16	$i$			&16					&128$\times$128$\times$16	&$\checkmark$&$\checkmark$\\
			Spatial Convolution										&8$\times$1$\times$3$\times$3			&16						&16 				&128$\times$128$\times$8	&$\checkmark$&$\checkmark$\\
			\textbf{3S Convolution (without Activation)}\\
			Spectral Convolution&128$\times$8$\times$1$\times$1	&128					&8					&128$\times$128$\times$16	&--&$\checkmark$\\
			Spatial Convolution&8$\times$1$\times$3$\times$3	&8						&8 					&128$\times$128$\times$16	&--&$\checkmark$\\
			Residual Addition				  &--	                    		&8+8=16				&8					&--							&--			 &$\checkmark$\\
			\hline	\hline
		\end{tabular}
	\label{tab:PZResNet_architecture}
\begin{center}
	\begin{tabular}{l c c c c c c}
			\hline	\hline
		Refinement Module		&Kernel shape 	  						& \# Input Channels 	&\# Output Channels &Output shape               & ReLU & ZM-norm \\
		\hline
		\textbf{Input }												&--										&31  					&31			 		&128$\times$128$\times$31	&--  		 &--	\\
		\textbf{3S Convolution-$j$ $j\in [1,2]$}	\\
		Spectral Convolution										&31$\times$31$\times$1$\times$1		&31				&31					&128$\times$128$\times$31	&$\checkmark$&--\\
		Spatial Convolution										&31$\times$1$\times$3$\times$3		&31				&31 				&128$\times$128$\times$31	&$\checkmark$&--\\
		\textbf{3S Convolution (without Activation)}\\
		Spectral Convolution										&31$\times$31$\times$1$\times$1		&31				&31					&128$\times$128$\times$31	&--			&--\\
		Spatial Convolution										&31$\times$31$\times$3$\times$3		&31				&31 				&128$\times$128$\times$31	&--			&--\\
		\textbf{Residual Addition}			  						&--	                    			&31+31=62		&31					&--							&--			&--\\
			\hline	\hline
	\end{tabular}
\end{center}
\label{tab:Refinement_mode}
\end{table*}

\subsubsection{Zero-mean normalization}
Our objective is to learn a zero-centric residual image.
However, non-linear activation layers (e.g., ReLU and Swish) involved in the residual learning module make the output feature maps to be non-negative, resulting in that their mean-values deviate from zero and likewise the estimated residual image. Besides, without additional constraints on the learned convolutional kernels, the convolution operation may also affect the mean-values of the output of each layer\footnote{Generally, the mean-value of a feature map will be preserved after a convolution operation only if the sum of the elements of the involved kernel is equal to 1}.

To achieve the objective, we propose a novel feature normalization process, namely \textit{Zero-Mean normalization} (\textit{ZM-norm}), which is performed on the spatial domain of each feature channel involved 
in the residual learning module. Specifically, in the forward propagation, the ZM-norm denoted by $\mathcal{Z}(\cdot)$ behaves as
\begin{align}
    \centering
    \mathcal{Z}(m_{b,k,c})&=m_{b,k,c}-\mathcal{E}(\mathbf{M})_{b,c}, \\
     \mathcal{E}(\mathbf{M})_{b,c}&=\frac{1}{HW}\sum_{k=1}^{HW}{m_{b,k,c}},\nonumber
\end{align}
where $m_{b,k,c}$ is the $(b, k, c)$-th entry of $\mathbf{M}\in\mathbb{R}^{B\times HW\times C}$, the input feature maps to a typical convolutional layer ($1\leq b\leq B$, $1\leq k\leq HW$, $1\leq c\leq C$). Here $b$, $k$, and $c$ indicate the mini-batch number, the spatial location, and  the channel number, respectively. The gradient of the training loss $\mathcal{L}$ can be back propagated through ZM-norm according to Eq. (\ref{backward}).
\begin{equation}
\begin{split}
\frac{\partial \mathcal{L}}{\partial m_{b,k,c}} &= \sum_{j=1}^{HW} \frac{\partial \mathcal{L}}{\partial \mathcal{Z}(m_{b,j,c})} \frac{\partial \mathcal{Z}(m_{b,j,c})}{\partial m_{b,k,c}}  \\
&= \frac{\partial \mathcal{L}}{\partial \mathcal{Z}(m_{b,k,c})} - \sum_{j=1}^{HW} \frac{\partial \mathcal{L}}{\partial \mathcal{Z}(m_{b,j,c})} \frac{\partial \mathcal{E}(\mathbf{M})_{b,c}}{\partial m_{b,j,c}}\\
&=\frac{\partial \mathcal{L}}{\partial \mathcal{Z}(m_{b,k,c})} -\frac{1}{HW}\sum_{j=1}^{HW}{\frac{\partial \mathcal{L}}{\partial \mathcal{Z}(m_{b,j,c})}}.\\
\end{split}
\label{backward}
\end{equation}

The proposed ZM-norm can be easily and efficiently implemented and integrated with existing deep learning architectures. Moreover, ZM-norm also introduces an additional advantage, i.e., \textit{it accelerates the training process}. The underlying reason is that during backpropagation, gradients are related to their corresponding feature values, and a zero-centric feature distribution could limit the gradient magnitude \cite{ioffe2015batch, lecun2012efficient}, therefore leading to more stable updates which speeds up convergence.

\textit{Remark}: We would like to point out that our ZM-norm is different from existing feature normalization methods, such as Layer Normalization \cite{ba2016layer}, Group Nomarlization \cite{wu2018group}, and Instance Normalization \cite{ulyanov2016instance}. Those normalization methods were mainly proposed to speed up the training process and improve the model generalization ability. Generally, they enforce the intermediate feature maps into a certain learnable distribution, through calculating 
the standard deviation and rescaling the feature magnitudes accordingly, which may cause loss of scale information \cite{lim2017enhanced}.
Our ZM-norm, however, is focused on eliminating the mean value of feature maps for regressing a zero-centric residual image that captures the high-frequency spatial details. 

\subsection{Refinement Module}
In the residual learning module, the residual image of each spectral band is \textit{independently} synthesized, which is then superimposed on the corresponding band of the up-sampled LR-HSI, leading to a coarse HR-HSI denoted by $\widehat{\mathbf{X}}\in\mathbb{R}^{HW\times S}$. However, the coherence among the bands of $\widehat{\mathbf{X}}$ cannot be well preserved.  Therefore, as shown in  Fig. \ref{fig:CRseNet}(c),  we further introduce a refinement module, in which all the spectral bands of $\widehat{\mathbf{X}}$ are simultaneously explored to enhance the reconstruction quality.  

The overall process of the refinement module $\mathcal{F}_r(\cdot)$ can be formulated as 
\begin{equation}
    \centering
    \widetilde{\mathbf{X}}=\mathcal{F}_r(\widehat{\mathbf{X}},\theta_r)+\widehat{\mathbf{X}},
\end{equation}
where $\theta_r$ is the weights to be learned, and $\widetilde{\mathbf{X}}\in\mathbb{R}^{HW\times S}$ is the finally reconstructed HR-HSI. More specifically, three 3S convolutional layers are employed to achieve feature extraction. Note that ZM-norm is no longer used in this module. The detailed implementation of this module is summarized in Table \ref{tab:PZResNet_architecture}. 
 
\subsection{Loss Function for Training}
Our PZRes-Net is end-to-end trained with the following loss function:
\begin{equation}
\mathcal{L}(\mathbf{X},\widetilde{\mathbf{X}}) = \frac{1}{HWS}\left(\left\|\mathcal{Z}(\mathbf{X})-\mathcal{Z}(\widehat{\mathbf{X}})\right\|_1+\lambda\left\|\mathbf{X}-\widetilde{\mathbf{X}}\right\|_1\right),
\label{eq:objective}
\end{equation}
where $\lambda>0$ is the parameter to balance the two terms, which is empirically set to 1, and $\|\cdot\|_1$ is the $\ell_1$-norm of a matrix, which returns the sum of the absolute of elements. The first term enforces PZRes-Net to learn the zero-centric residual image, while the second term encourages the reconstructed HR-HSI to be close to the ground-truth one in sense of the mean absolute error.
\begin{figure*}
    \centering
    \includegraphics[width=\textwidth,trim=0 0 0 5,clip]{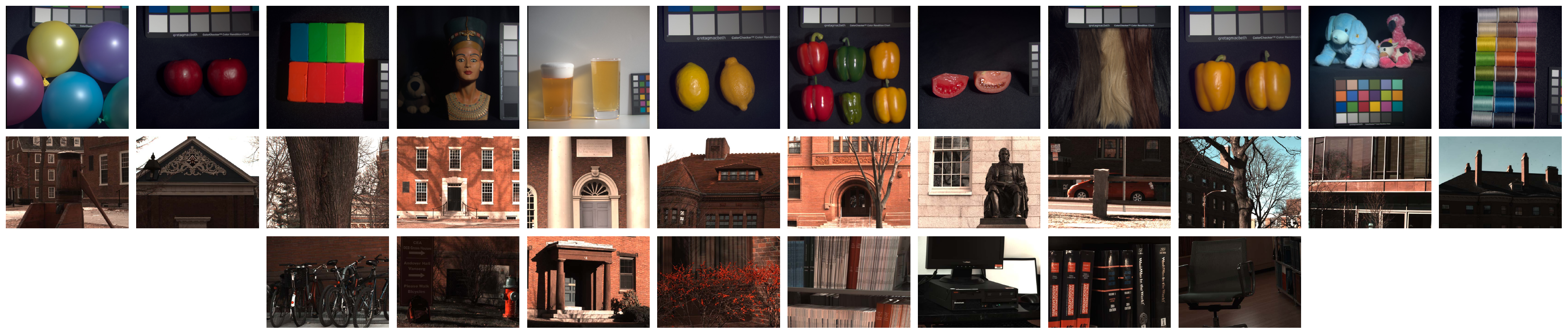}
    \caption{Illustration of the testing images  from CAVE (the 12 images in the 1$^{st}$ row) and HARVARD (the 20 images in the 2$^{nd}$ and 3$^{rd}$ rows) in the evaluation on synthetic data.}
    \label{fig:testing_image}
\end{figure*}

\section{Experiments}
\label{sec:EXP}

\subsection{Experiment Settings}

\subsubsection{Implementation details} 
We adopted ADAM \cite{kingma2014adam} optimizer with the exponential decay rates  $\beta_1 = 0.9$ and $\beta_2 = 0.999$ for the first and second moment estimates, respectively. 
The learning rate of our PZRes-Net was initialized as $1e-3$ and the cosine annealing decay strategy was employed to decrease it gradually, ended with $1e-5$. During training, we kept the same number of iterations to be 32000. We implemented the model with PyTorch, and the batch size was set to 10 for CAVE and 30 for HARVARD. All the experiments were conducted on Linux 18.04 with Intel Xeon E5-2360 CPU and NVIDIA 2080TI GPUs. The code will be publicly available.  Table \ref{tab:PZResNet_architecture} summarizes the implementation details of our network architecture.
\subsubsection{Compared Methods}
We compared our PZRes-Net with 8 state-of-the-art HSI super-resolution approaches, including 4 traditional methods, i.e., hyperspectral super-resolution (HySure) \cite{simoes2014convex}, nonnegative structured sparse representation (NSSR) \cite{dong2016hyperspectral}, the clustering manifold structure-based method (CMS) \cite{zhang2018exploiting}, and the low tensor-train rank-based method (LTTR) \cite{dian2019learning}, and 4 DNN-based methods, i.e., cross-attention in coupled unmixing nets (CUCaNets) \cite{yao2020cross}, deep HSI sharpening (DHSIS) \cite{dian2018deep}, multispectral and hyperspectral image fusion network (MHF) \cite{xie2019multispectral}, and deep blind HSI fusion network (DBIN+) \cite{wang2019deep}. For fair comparisons, the same data pre-processing was implemented in all methods, the DNN-based methods under comparison were trained with the codes provided by the authors with suggested parameters over the same training data as ours, and the same protocol in \cite{dian2019learning, wang2019deep} was used to evaluate the experimental results of all methods.  

\subsubsection{Quantitative Metrics}
For a comprehensive evaluation, we adopted four commonly-used quantitative metrics:
\begin{itemize}
    \item Peak Signal-to-Noise Ratio (PSNR):
\begin{equation}
\text{PSNR}(\mathbf{X},\widetilde{\mathbf{X}}) = -\frac{10}{S} \sum_{k=1}^{S} \log(\text{MSE}(\mathbf{X}_k,\widetilde{\mathbf{X}}_k)),
\end{equation}
where $\widetilde{\mathbf{X}}_k\in\mathbb{R}^{H\times W}$ and $\mathbf{X}_k\in\mathbb{R}^{H\times W}$  are the $k$-th ($1\leq k\leq S$) spectral bands of $\widetilde{\mathbf{X}}$ and $\mathbf{X}$, respectively, and $\text{MSE}(\cdot)$ returns the mean squares error between the inputs. 
\item  Average Structural Similarity Index (ASSIM):
\begin{equation}
\text{ASSIM}(\mathbf{X},\widetilde{\mathbf{X}}) = \frac{1}{S} \sum_{k=1}^{S}\text{SSIM}(\mathbf{X}_k,\widetilde{\mathbf{X}}_k),
\end{equation}
where $\text{SSIM}(\cdot,\cdot)$ \cite{wang2004image} computes the SSIM value of a typical spectral band.

\begin{figure*}
\centering
\includegraphics[width=0.80\linewidth]{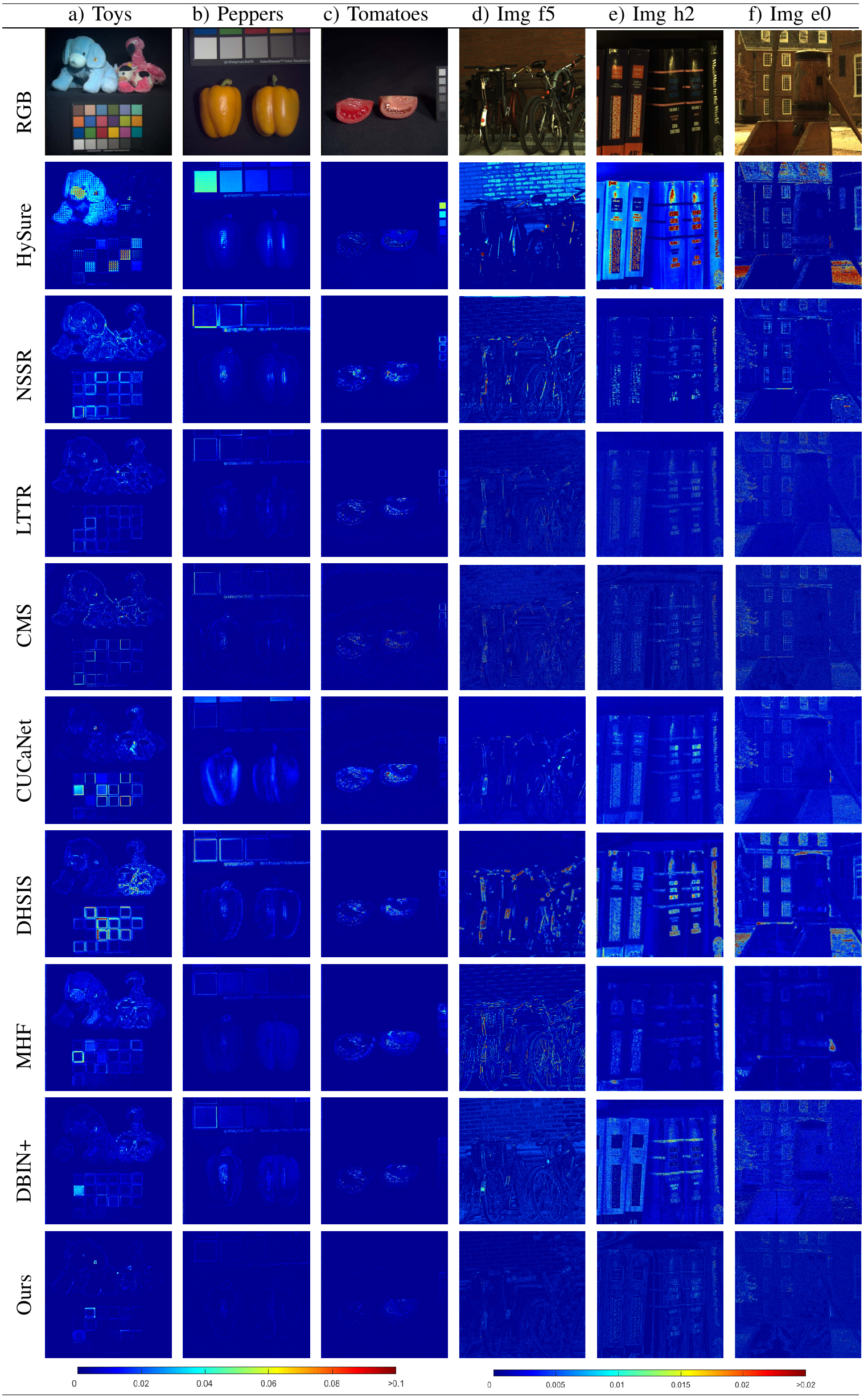}
\caption{Comparisons of the error maps between the spectral bands of reconstructed HR-HSIs by different methods and the corresponding ground-truth ones. (a)-(c): spectral bands of images from the CAVE dataset at wavelength 600 nm; (d) the spectral band of the image from the HARVARD dataset at wavelength 420 nm; (e)-(f): spectral bands of images from the HARVARD dataset at wavelength 520 nm.}
\label{fig:final_result_0}
\end{figure*}
\item Spectral Angle Mapper (SAM):
\begin{equation}
\text{SAM}(\mathbf{X},\widetilde{\mathbf{X}}) = \frac{1}{HW} \sum_{j=1}^{HW}\arccos{\left(\frac{\widetilde{\mathbf{x}}^\textsf{T}_j\mathbf{x}_j}{\|\widetilde{\mathbf{x}}_j\|_2\|\mathbf{x}_j\|_2}\right)},
\end{equation}
where $\widetilde{\mathbf{x}}_j\in \mathbb{R}^{S}$ and $\mathbf{x}_j\in \mathbb{R}^{S}$ are the  spectral signatures of the $j$-th ($1\leq j\leq HW$) pixels of $\widetilde{\mathbf{X}}$ and $\mathbf{X}$, respectively, and $\|\cdot\|_2$ is $\ell_2$ norm of a vector.  
\item
Erreur Relative Global Adimensionnelle Synthese (ERGAS):
\begin{equation}
\text{ERGAS}(\mathbf{X},\widetilde{\mathbf{X}}) = \frac{100}{r} \sqrt{\frac{1}{S}\sum_{k=1}^{S}\frac{\text{MSE}(\mathbf{X}_i,\widetilde{\mathbf{X}}_i)}{\mu^2_{\widetilde{\mathbf{X}}_i}}},
\end{equation}
where $\widetilde{\mu}_k$ is the mean-value of the $k$-th spectral band of $\widetilde{\mathbf{X}}$, and $r$ is the scale factor.
\end{itemize}

\begin{figure*}
\centering
\subfigure[Toys]{\includegraphics[width=0.15\linewidth]{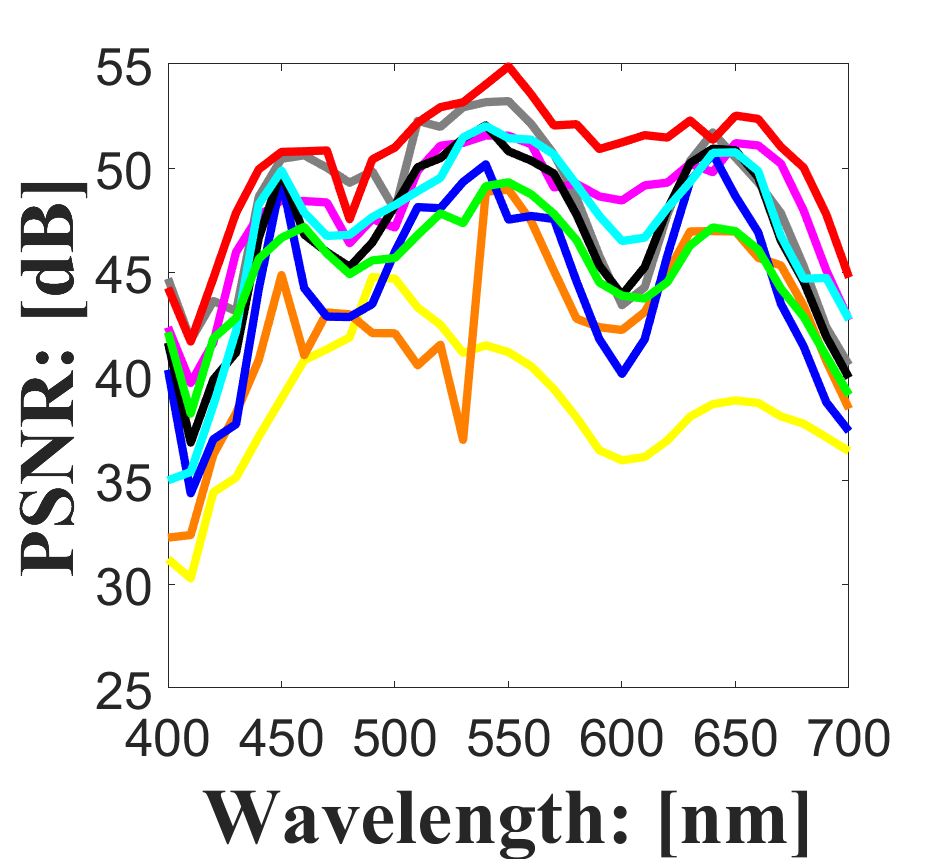}}
\subfigure[Peppers]{\includegraphics[width=0.15\linewidth]{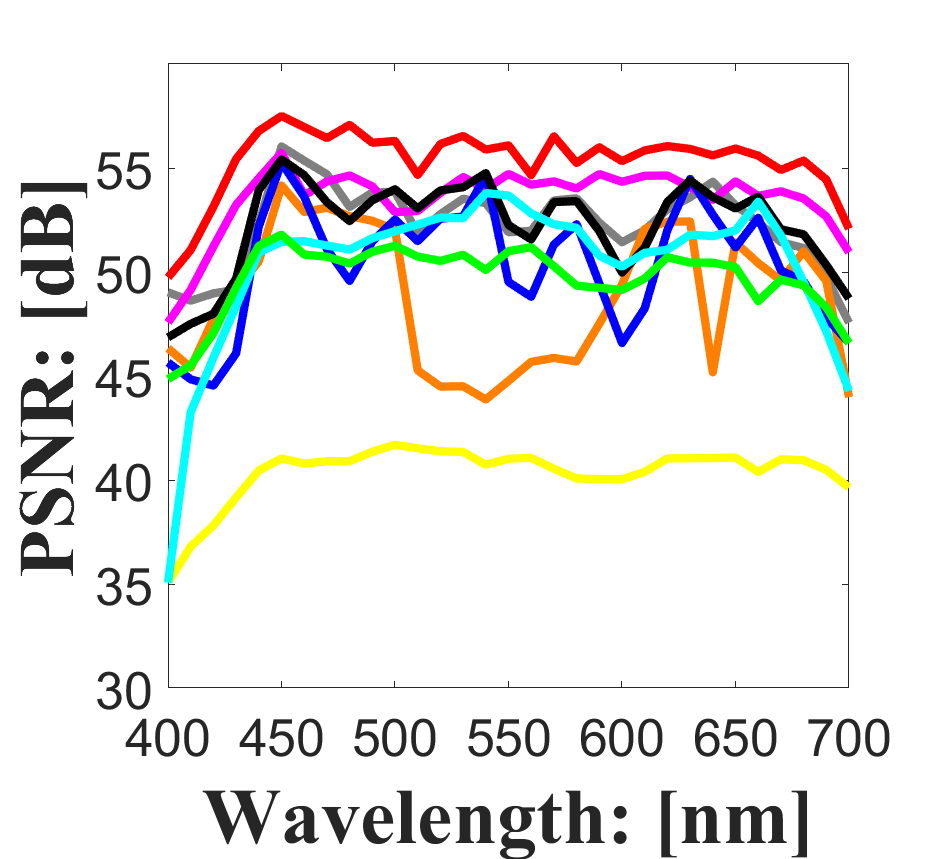}}
\subfigure[Tomatoes]{\includegraphics[width=0.147\linewidth]{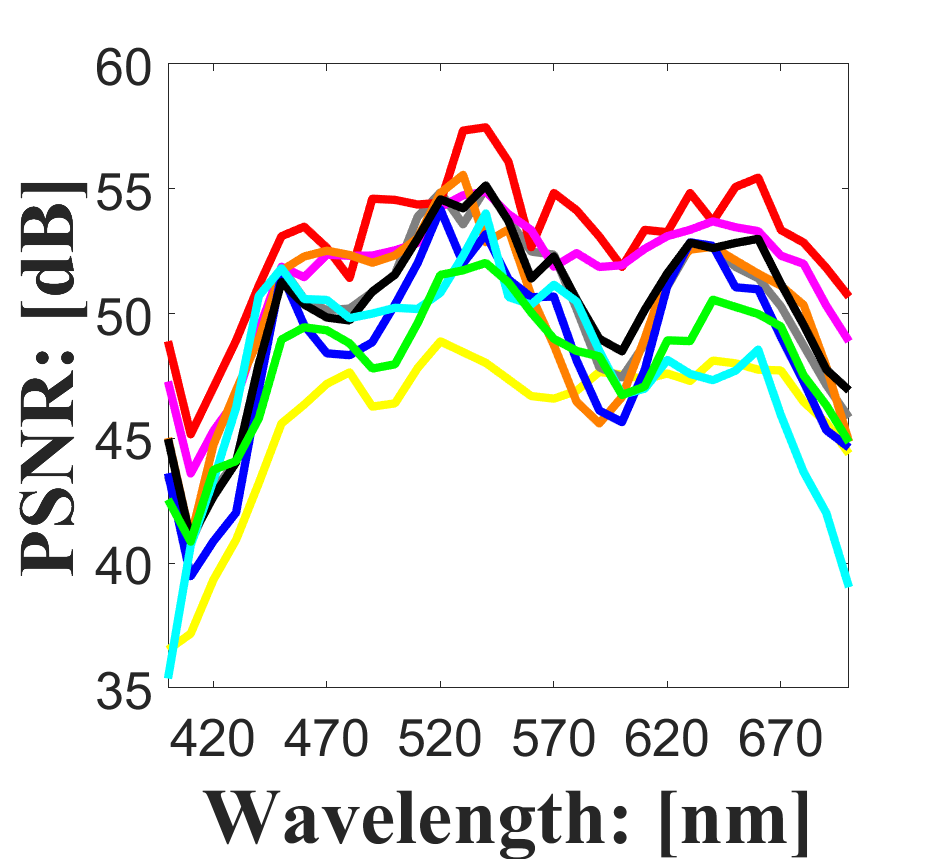}}
\subfigure[Img f5]{\includegraphics[width=0.15\linewidth]{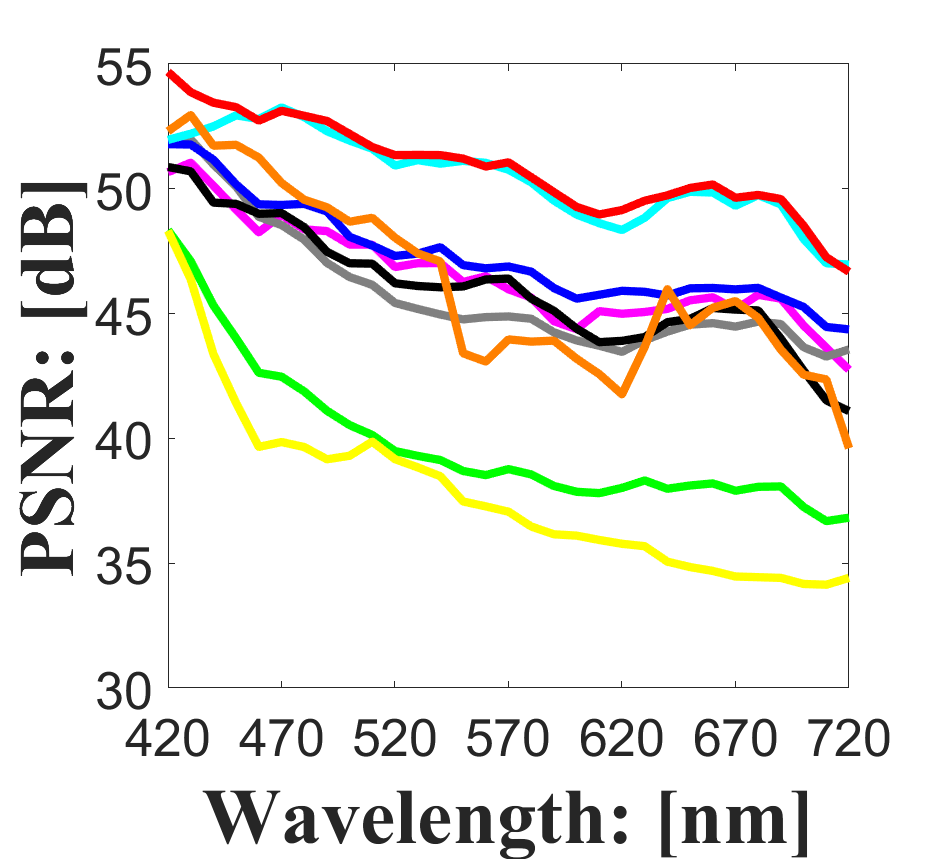}}
\subfigure[Img h2]{\includegraphics[width=0.15\linewidth]{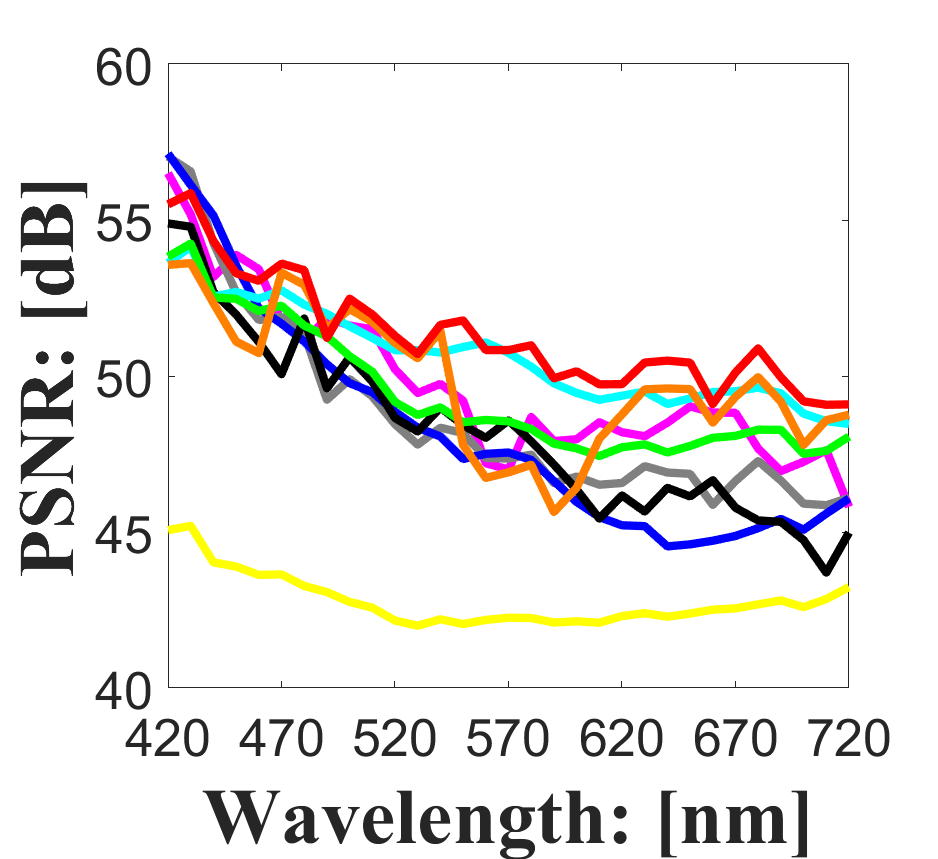}}
\subfigure[Img e0]{\includegraphics[width=0.15\linewidth]{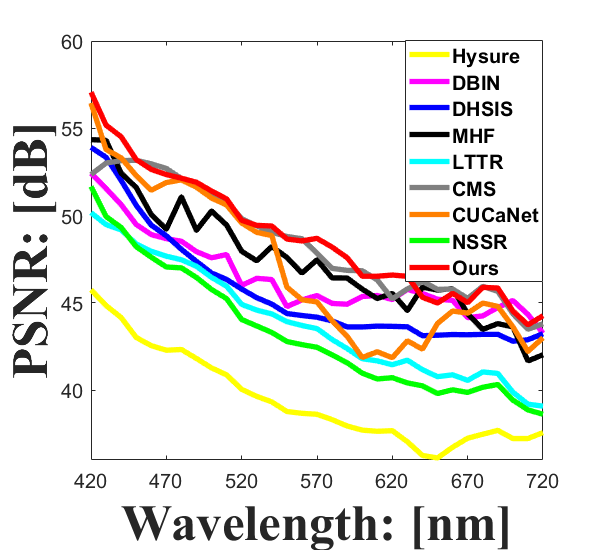}}
\caption{Quantitative comparisons of the proposed PZRes-Net with state-of-the-art methods in terms of the PSNR value of each spectral band of the reconstructed HR-HSI. Note that these 6 results correspond to the 6 images illustrated in Fig. \ref{fig:final_result_0}.}
\label{fig:PSNR_Curves}
\end{figure*}
In addition to the above-mentioned quantitative metrics for evaluating the output image quality, we also added up the number of neural network parameters (\# Param)  and the number of floating number operations per-inference (FLOPs) of the four DNN-based methods to compare their efficiency. Note that the unsupervised CUCaNet was trained on each testing sample independently. For a fair comparison, we multiplied the FLOPs of a single inference with the number of training epochs $N$, which was set to 10000 in our experiments.
\begin{table*}
	\caption{Quantitative comparisons of different methods in terms of 4 metrics over the 12 and 20 testing samples from the CAVE and HARVARD datasets, respectively. For PSNR and ASSIM, the larger, the better. For SAM and ERGAS, the smaller, the better. Note that ERGAS can be compared only under the same scale. For FLOP and \# parameters, the smaller, the more efficient. $N$ denotes the number of training epochs for CUCaNet. The top four are traditional methods, and the bottom four as well as ours are DNN-based. The best results are bold.}
	\label{tab:synthetic}
		\resizebox{\linewidth}{!}{
		\begin{tabular}{lcc|ccccc|ccccc}
			\hline	\hline
			&&&\multicolumn{5}{c|}{CAVE} & \multicolumn{5}{c}{HARVARD}\\
			\hline
			Methods	&Scale& \# Pram & FLOPs& PSNR & ASSIM &SAM &ERGAS  & FLOPs & PSNR & ASSIM & SAM & ERGAS\\
			\hline	
			NSSR \cite{dong2016hyperspectral}		&$\times$8&--	 &--	&43.82  &0.987	&4.07	&0.84 &--       &45.92  &0.980	&3.46&1.20  \\
			HySure \cite{simoes2014convex}		    &$\times$8&--	 &--	&37.35  &0.945	&9.87	&2.01 &--     &43.88  &0.975	&4.20&1.56   \\
			LTTR \cite{dian2019learning}			&$\times$8&--    &--	&46.35  &0.990	&3.76	&0.80 &--	    &46.45  &0.982	&3.46&1.20  \\
			CMS \cite{zhang2018exploiting} &$\times$8&--&-- &47.32&0.997&3.07   &0.81  &--  &46.52  &0.988 &3.04 &1.43   \\
			DHSIS \cite{dian2018deep}				&$\times$8&0.5M	 &270G  &45.70 	&0.990	&3.95 	&0.75 &1.17T &46.42   &0.982  &3.54 &1.17   \\
			MHF \cite{xie2019multispectral}			&$\times$8&1.3M	 &476G  &46.81	&0.992	&3.83   &0.75 & 2.04T  &46.40   &0.982  &3.45&1.20  \\
			CUCaNet \cite{yao2020cross}&$\times$8&1.0M &153G $\times N$&45.21 &0.991 &4.28 &1.36 &0.61T$\times N$ &45.90&0.980 &3.51 &1.55\\
			DBIN+ \cite{wang2019deep}				&$\times$8&0.7M	 &4,095G&47.65	&0.992	&3.12	&0.58 & 17.58T &46.67  &0.983  &3.42&1.15   \\
			Ours 									&$\times$8&0.7M  &271G  &\textbf{50.94}  &\textbf{0.998} &\textbf{2.63}	&\textbf{0.43} & 1.08T  & \textbf{47.52}&\textbf{0.989}&\textbf{2.83}&\textbf{1.07} \\
			\hline	\hline
			Methods	&Scale&\#Pram& FLOPs& PSNR & SSIM &SAM &ERGAS & FLOPs & PSNR & ASSIM & SAM & ERGAS \\
			\hline
			NSSR \cite{dong2016hyperspectral}		&$\times$32&--		&--	  &39.89&0.958	&8.33	&0.63 & --	& 	40.17&0.956&5.13&0.3503 \\
			HySure \cite{simoes2014convex}			&$\times$32&--		&--	  &30.14&0.913	&10.35	&2.35 &-- & 	35.17&0.937&7.15&0.5041 \\
			LTTR \cite{dian2019learning}			&$\times$32&-- 		&--	  &41.03&0.970	&8.29	&0.39&--	&	43.25&0.975&5.45&0.3062 \\
			CMS \cite{zhang2018exploiting}		    &$\times$32 & --	    &--   &42.97  &0.989	   &7.07 &0.25   &-- &44.59&0.981  &4.36 &0.4790 \\
			DHSIS \cite{dian2018deep}				&$\times$32&0.5M	&270G &40.51&0.967	&8.01 	&0.41 & 1.17T	&	43.73&0.978&5.31&0.3107 \\
			MHF \cite{xie2019multispectral}			&$\times$32&1.7M	&673G &40.12&0.962	&7.30   &0.45 & 2.46T	&	44.31&0.981&4.90&0.2955  \\
			CUCaNet \cite{yao2020cross}&$\times$32&0.9M &152G $\times N$ & 41.41 &0.960 &7.10 &0.40 &0.61T$\times N$ &42.29&0.970 &5.91 &0.4013 \\
			DBIN+ \cite{wang2019deep}				&$\times$32&1.67M	&4,571G&42.83&0.988	&6.63	&0.27 & 31.05T&	44.24&0.980&5.32&0.2964  \\
			Ours 									&$\times$32&0.7M 	&271G  &\textbf{45.77}&\textbf{0.991}	&\textbf{6.25}	&\textbf{0.23} & 1.08T	&\textbf{45.47}&\textbf{0.986}&\textbf{4.19}&\textbf{0.2878} \\
			\hline	\hline
		\end{tabular}
		}
\end{table*}

\subsection{Evaluation on Synthetic Data}
In this scenario, two commonly used benchmark HSI datasets, i.e., CAVE \footnote{http://www.cs.columbia.edu/CAVE/databases/} \cite{yasuma2010generalized} and HARVARD\footnote{http://vision.seas.harvard.edu/hyperspec/} \cite{chakrabarti2011statistics}, were used to generate \textit{synthetic} hybrid inputs. Specifically, the CAVE dataset contains 32 indoor HSIs of spatial dimensions 512 $\times$ 512 and spectral bands 31 captured by a generalized assorted pixel camera with an interval wavelength of 10nm in the range of 400-700nm. Following \cite{xie2019multispectral,xie2020mhf}, we randomly selected 20 images for training and the remaining 12 images for testing. The HARVARD dataset contains 50 indoor and outdoor HSIs recorded under the daylight illumination, and 27 images under the artificial or mixed illumination. Each HSI consists of 31 spectral bands of spatial dimensions 1024$\times$1392, whose wavelengths range from 420 to 720 nm. Following \cite{dian2018deep, wang2019deep},  only the 50 daylight illumination images were used in our experiments. Moreover, the first 30 HSIs are used for training, and the last 20 ones for testing. Fig. \ref{fig:testing_image} shows all the testing images used in the evaluation on synthetic data (the remaining ones are training data). Following \cite{wang2019deep, dian2019learning, xie2019multispectral}, all the LR-HSIs (with down-sampling scale $r$) used in this scenario were acquired through following two steps: (1) an $r \times r$ Gaussian kernel is used to blur HSIs; and (2) the blurred HSIs were regularly decimated every $r$ pixels in the spatial domain. We simulated the HR-MSI (RGB image) of the same scene by integrating the spectral bands of an HSI with the widely used response function of the Nikon D700 camera\footnote{http://www.maxmax.com/spectral\_response.htm}. All the DNN-based methods were trained separately on the CAVE and HARVARD datasets.

\subsubsection{Results on the CAVE dataset}

\begin{figure*}
\includegraphics[width=\textwidth,trim=0 0 0 5,clip]{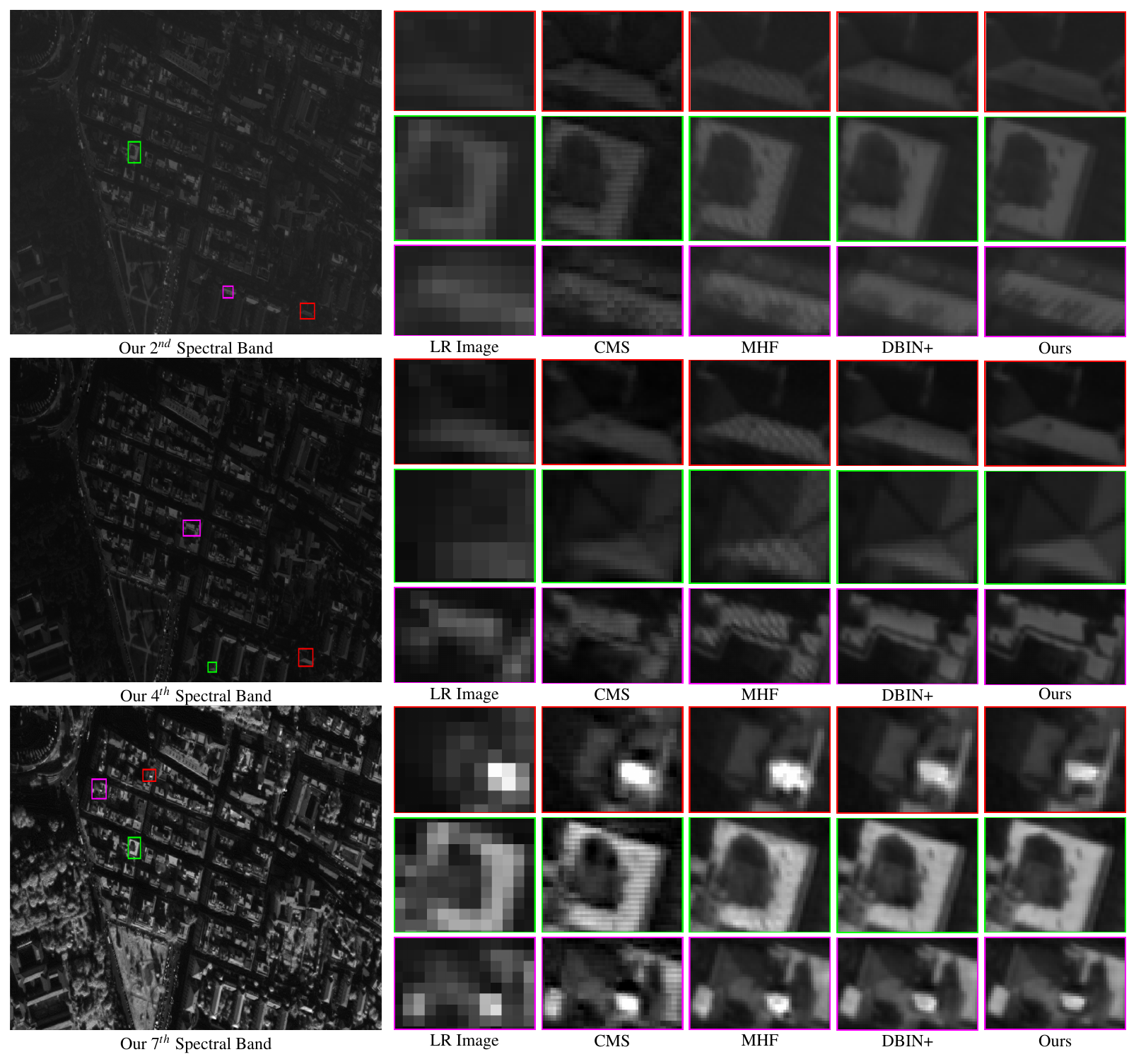}
\caption{Experimental results on the real dataset, WV-2, and the visualized image corresponds to the bottom left part of WV-2. The scale factor is equal to 4. We zoomed in the selected regions within the colored boxes  with the  'nearest' interpolation 5 times for a better comparison.}
\label{fig:Real_world}
\end{figure*}

\begin{figure*}
\includegraphics[width=\textwidth,trim=0 0 0 5,clip]{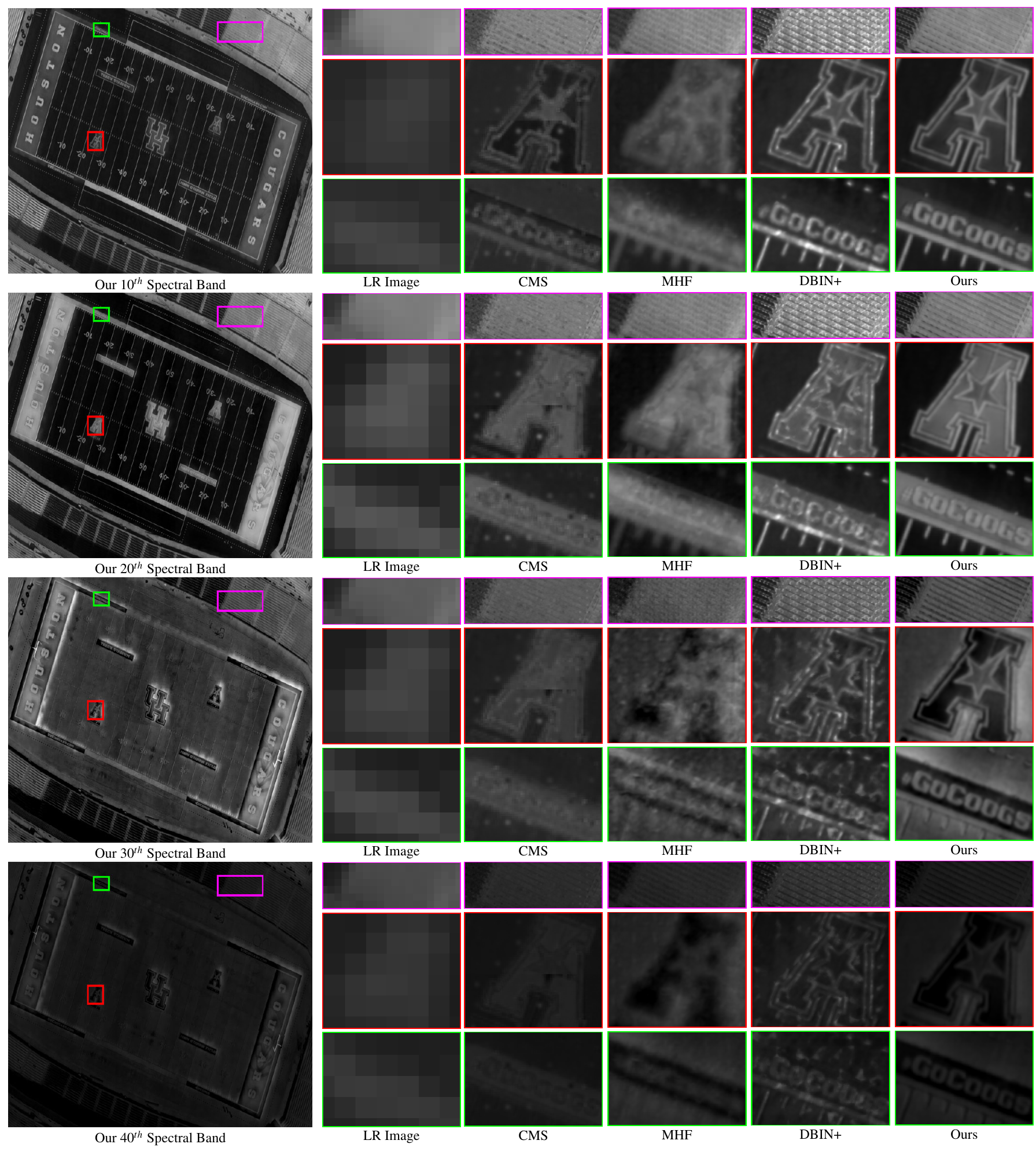}
\caption{Experimental results on the real dataset, NCALM, and the visualized image corresponds to a patch from the right part of NCALM. The scale factor is equal to 20. We zoomed in three selected regions within the colored boxes with the 'nearest' interpolation 5 times for a better comparison.}
\label{fig:Real_world2}
\end{figure*}
As listed in the left side of Table \ref{tab:synthetic}, we can see that the proposed method with only \textit{0.7M} parameters consistently surpasses all the  methods under comparison significantly in terms of all the four metrics under both up-sampling scales. Although the parameter sharing strategy in DBIN+ could help to save parameters to some extent, its iterative refinement manner costs much more computation resources as demonstrated by the high FLOPs value. To be specific, DBIN+ consumes the FLOPs of 4095G, which is 15$\times$ of that of the proposed method, and thus its utilization in practice may be severely restricted. The traditional methods, e.g. NSSR, LTTR, show good reconstruction performance in the $8\times$ reconstruction. However, as the up-sampling scale rises to 32, the matrix/tensor factorization-based method LTTR show a sharp deterioration, which is caused probably by the model's limited representation ability or by the failure of prior knowledge. Through exploring the clustering structure in HSIs, CMS outperforms the other traditional methods and even beats all the compared  DNN-based methods in the $32\times$ experiment. When the performance of all the compared methods drops for a larger up-sampling scale, the proposed PZRes-Net model still maintains the highest performance in terms of the 4 metrics.
\subsubsection{Results on the HARVARD dataset}
The experimental results on the HARVARD dataset of different methods are listed in right side of Table \ref{tab:synthetic}. Note that the HARVARD dataset is more challenging than the CAVE dataset due to the higher resolution and more complex scenario. The significant superiority of our method over state-of-the-art methods is further validated. Specifically, CMS, MHF, and DBIN+ produce comparable performance, which exceeds other methods not significantly. However, our PZRes-Net with PSNR of 47.52 dB (resp. 45.47 dB) and SAM of 2.83 (4.19) greatly pushes forward the limits at the scale of 8$\times$ (resp. 32$\times$). Moreover, compared with the CAVE dataset, the SAM values of all methods on the HARVARD dataset are smaller, which may be caused by the dark background in the CAVE dataset. With the same absolute error, the pixels with low intensities are easy to output high SAM values. 
\subsubsection{Visual comparisons} 
The error maps between the spectral bands of reconstructed HR-HSI by different methods and the ground-truth ones are shown in Fig. \ref{fig:final_result_0}. Accordingly, Fig. \ref{fig:PSNR_Curves} provides the PSNR values of each spectral band of the visualized images in Fig. \ref{fig:final_result_0} for reference. For the images from the CAVE dataset, i.e., Figs. \ref{fig:final_result_0}(a)-(c), the traditional methods, e.g. NSSR, LTTR,  and CMS, have comparable performance to the two state-of-the-art DNN-based methods, i.e.,  DBIN+ and MHF, but the shapes of objects can be easily inferred from the error maps of all the compared methods due to the large errors at the boundaries.
However, our PZRes-Net consistently produces the smallest errors over all the images at  a low computational cost, and there are only subtle errors, which are nearly invisible.
Similar observations can be obtained for the images from the HARVARD database, i.e., Figs. \ref{fig:final_result_0}. (d)-(f). The results convincingly demonstrate the advantage of our method. 

\subsection{Evaluation on Real Data}
In this scenario, we evaluated our PZRes-Net over two real datasets named WorldView-2\footnote{https://www.harrisgeospatial.com/Data-Imagery/Satellite-Imagery/High-Resolution/WorldView-2} (WV-2)  and National Center for Airborne Laser Mapping\footnote{http://www.grss-ieee.org/community/technical-committees/data-fusion/2018-ieee-grss-data-fusion-contest/} (NCALM). WV-2 has been widely adopted in previous works \cite{xie2020mhf, xie2019multispectral,wang2019deep,nikolakopoulos2015quality,wu2019remote,alimuddin2012assessment}, which contains an 8-band HSI of spatial dimensions $418 \times 685$ and a 3-band MSI of spatial dimensions $1677 \times 2633$, and the data were captured on a commercial satellite. NCALM provided by the 2018 IEEE GRSS Data Fusion Contest \cite{xu2019advanced} contains a 3-band MSI of spatial dimensions $24040 \times 83440$, which was captured with the DiMAC ULTRALIGHT+ with the focal length equal to 70mm,  and a 48-band HSI of spatial dimensions $1202 \times 4172$ covering the spectral range from 380 to 1050 nm, which was captured with the ITRES CASI 1500. The sensors were placed on a Piper PA-31- 350 Navajo Chieftain aircraft. As the ground-truth data are not available, following \cite{xie2019multispectral}, we generated the training data in the following way. Specifically, for the WV-2 dataset, we first downsampled the 3-band RGB and 8-band MSI images leading to an Hr-MSI of spatial dimensions $416 \times 656$ and an Lr-HSI of spatial dimensions $104 \times 164$, and then trained all models equally on the top half part of the resulting HR-MSI and LR-HSI. During training, the corresponding part of the original HSI was used as supervision. The bottom half part of the original data excluded in the training process was used for testing. For the NCALM dataset, as the huge spatial dimensions of the original data cause a memory issue, we first extracted an HSI patch of spatial dimensions $1200 \times 600$ and an MSI patch of spatial dimensions $24000 \times 12000$ from the rightmost of the original data, which were then spatially downsampled to $60 \times 30$ and $1200 \times 600$ to form the training data. We then selected a patch from the remaining left part of the original data for testing. Here we only show the results of the three methods, i.e., DBIN+, MHF, and CMS, which always achieve the top performance among all the compared methods in the previous scenario, due to the space limitation. 

The three sub-images in Fig. \ref{fig:Real_world} visualize three spectral bands of the reconstructed HR-HSI of different methods, where it can be observed that in the $4^{th}$ spectral band, our PZRes-Net is able to reconstruct both high-frequency details (eaves in the magenta box of the $2^{nd}$ spectral band) and smooth parts (the roof in the magenta and green boxes of the $4^{th}$ and $7^{th}$ spectral bands, respectively) very well. However, both DBIN+ and MHF fail to reconstruct high-frequency spatial details, resulting in blurring boundaries, and unexpected visual artifacts also appear in the smooth regions. The non-deep learning-based method CMS produces indistinct results, which may be caused by the misalignment due to the limited model representation learning ability. Fig. \ref{fig:Real_world2} illustrates the results of different methods on the NCALM dataset. Due to the large scale factor, i.e., $20\times$, the performance of the three compared methods decreases significantly; however, the proposed method still works well and produces much better results. All these visually pleasing results of our method are credited to its advantage on modeling the cross-modality information.

\subsection{Ablation Study}
We carried out extensive ablation studies to comprehensively analyze the three key components of our PZRes-Net model over the CAVE dataset.

\begin{table}
		\caption{Ablation studies towards the refinement module and the residual-dense architecture. The best results are bold.}
		\label{Refinement_module}
		\centering
		\begin{tabular}{l|cccc}
			\hline	\hline
			Models & PSNR & ASSIM &SAM &ERGAS \\
			\hline
			w/o Refinement 				&50.56 &0.994&3.07 &0.47\\
			\multirow{2}{0.3\linewidth}{w/o residual-dense aggregation} 	&\multirow{2}{*}{50.45} &\multirow{2}{*}{0.995}&\multirow{2}{*}{2.95} &\multirow{2}{*}{0.47}\\\\
			Full PZRes-Net						&\textbf{50.94}&\textbf{0.998}&\textbf{2.63} &\textbf{0.43}\\
			\hline	\hline
		\end{tabular}
\end{table}

\begin{table}[t]
\centering
		\caption{Comparisons of our PZRes-Net with different numbers of stages in the residual learning module. The best results are bold.}
		\label{tab:Progressive_manner}
		\begin{tabular}{c|cccc}
			\hline	\hline
			Models & PSNR & SSIM &SAM &ERGAS \\
			\hline
			One stage 	&50.47 &0.994&2.83 &0.47\\
			Two stages	&50.82 &0.995&2.71 &0.45\\
			Three stages&\textbf{50.94}&\textbf{0.998}&\textbf{2.63} &\textbf{0.43}\\
			\hline	\hline
		\end{tabular}
\end{table}
\begin{table}[t]
		\caption{Comparisons of our PZRes-Net equipped with different kinds of convolutional layers for feature extraction. The best results are bold.}
		\label{convolutional_layer}
		\centering
		\resizebox{\linewidth}{!}{
		\begin{tabular}{l|cccccc}
			\hline	\hline
			Models & \# Param & FLOPs & PSNR & ASSIM &SAM &ERGAS \\
			\hline
			2-D conv.	&1.4M	&292G		&48.77 &0.990	&3.10 	&0.55\\
			3-D conv.	&0.5M	&565G 		&50.59 &0.997	&2.71 	&0.44\\
			3S conv.	&0.7M	&271G 		&\textbf{50.94} &\textbf{0.998} 	&\textbf{2.63} 	&\textbf{0.43}\\
			\hline	\hline
		\end{tabular}}
\end{table}

\begin{table}[t]
\centering
	\caption{Comparisons of the proposed method equipped with kernels of different sizes in the spatial convolution of the 3S convolutional layer.}
	\label{tab:KZ}
		\begin{tabular}{l|cccccc}
		\hline	\hline
		Kernel size & Params & FLOPs & PSNR & ASSIM &SAM &ERGAS \\
		\hline
		$1\times 1$ &0.3M&70G&50.15	  &0.995 &2.80  &0.49\\
		$3\times 3$	&0.7M&271G&50.94    &0.998 &2.63  &0.43\\
		$5\times 5$	&1.2M&446G&51.08	  &0.998 &2.60  &0.42\\
		\hline	\hline
	\end{tabular}
\end{table}

\begin{table}[t]
\centering
	\caption{Illustrations of the necessarily of the mean-value invariant up-sampling and ZM-norm in our framework. A restriction-free transposed convolutional layer is learned w/o the mean-value invariant property for up-sampling when the marker is ``$\times$" under the $1^{st}$ column. ZM-norm was not applied when the marker is ``$\times$" under the $2^{nd}$ column. The best results are bold.
	}
	\label{normalization}
		\resizebox{0.95\linewidth}{!}{
		\begin{tabular}{cc|cccc}
		\hline	\hline
	 \multirow{2}{0.3\linewidth}{Mean-value invariant up-sampling} & \multirow{2}{*}{ZM-norm} & \multirow{2}{*}{PSNR} & \multirow{2}{*}{ASSIM} &\multirow{2}{*}{SAM} &\multirow{2}{*}{ERGAS} \\ \\
		\hline	
		$\times$ & $\times$		&47.60	 &0.992 &3.10  &0.77\\
		$\times$ & $\checkmark$  	&47.45	 &0.992 &3.14  &0.77\\
		$\checkmark$ & $\times$	&46.42	&0.989	&3.75 	&0.81  \\
		$\checkmark$ & $\checkmark$	&\textbf{50.94}  &\textbf{0.998}	&\textbf{2.63}	&\textbf{0.43}  \\
		\hline	\hline
	\end{tabular}}
\end{table}

\begin{table}[t]
\centering
	\caption{Investigations of the up-sampling process on reconstruction quality. The best results are bold.}
	\label{interpolation}
		\begin{tabular}{l|cccc}
		\hline	\hline
		Models & PSNR & ASSIM &SAM &ERGAS \\
		\hline
		Learned up-sampling    	&47.45	 &0.992 &3.14  &0.77\\
		Bi-cubic			 &50.79	 &0.997 &2.93  &0.72\\
		Bi-linear			 &\textbf{50.94}  &\textbf{0.998}	&\textbf{2.63}  &\textbf{0.43}\\
		\hline	\hline
	\end{tabular}
\end{table}

\subsubsection{The refinement module}
As aforementioned, we used a refinement module to boost the reconstruction performance by simultaneously exploring the coherence among all spectral bands of the resulting coarse HR-HSI. We trained PZRes-Net without (w/o) such a module. By comparing the $1^{st}$ and the $3^{rd}$ rows of Table \ref{Refinement_module}, the effectiveness of this module can be validated. 

\subsubsection{The residual-dense architecture}
To facilitate feature extraction, residual dense aggregation is embedded in our network.  Here we investigated the contribution of such an architecture by training our PZRes-Net without the all the identity mapping and dense connections of each stage. Additionally, we also widened the network such that the modified network has approximately the same number of parameters for a fair comparison, i.e., for the first stage we increased the widths of the three stages from 16, 32, 62 to 24, 48, and 93, respectively. As shown in Table \ref{Refinement_module}, compared with full PZRes-Net, the PSNR and ASSIM values of PZRes-Net w/o residual-dense aggregation decreases about 0.5 dB and 0.003, respectively, and the SAM and ERGAS values increases 0.32 and 0.04, respectively, validating the advantage of the residual-dense architecture. 

\subsubsection{The progressive spectral embedding scheme}
Inspired by the progressive strategy in image super-resolution, in our framework the spectral information of the input HSI is progressively fed  into the network to reconstruct HR-HSI. In order to validate the advantage of such a  progressive manner, we trained our PZRes-Net with different numbers of stages. ``One stage" means all the spectral bands of the up-sampled LR-HSI are fed into the network at the beginning. Note that we kept the modified models under different settings with the same number of parameters though varying the width of the networks for fair comparisons. As shown in Table \ref{tab:Progressive_manner}, the reconstruction quality gradually improves with the increasing the number of stages, and the growth rate is getting smaller. Thus, the advantage of the proposed progressive embedding scheme is convincing validated. Based on this observation, we use 3 stages in our framework.

\subsubsection{The 3S convolution}
In our PZRes-Net, 3S convolution enables efficient HSI processing. To investigate its efficiency and effectiveness, we trained our model by respectively replacing the 3S convolutional layers with 2-D and 3-D convolutional layers 
while keeping approximately the same amount of parameters or FLOPs. The experimental results are listed in Table \ref{convolutional_layer}. From Table \ref{convolutional_layer}, it can be seen that when using 2-D convolution, the performance drops sharply from 50.94 dB to 48.77 dB because 2-D convolution has a very limited ability to capture spectral information. Although 3-D convolution is able to achieve good performance, it consumes much more computation resources. Moreover, it is quite time-consuming. Our PZRes-Net with 3S convolution is capable of well balancing the efficiency and effectiveness. 
Moreover, we investigated how the kernel size of the spatial convolution involved in the proposed 3S convolutional layer affects the overall performance. Specifically, we varied the kernel size in the range of $1 \times 1$, $3\times 3$, and $5 \times 5$. The results are listed in Table \ref{tab:KZ}, where it can be seen that a larger kernel size is able to produce better reconstruction quality, but accordingly the computational complexity increases dramatically. Considering the trade-off between reconstruction accuracy and computational complexity, the kernel of size $3\times 3$ is the best choice.
\begin{figure}[t]
	\centering
	\subfigure[Testing PSNR]{\includegraphics[width=0.43\linewidth,trim=0 0 0 0,clip]{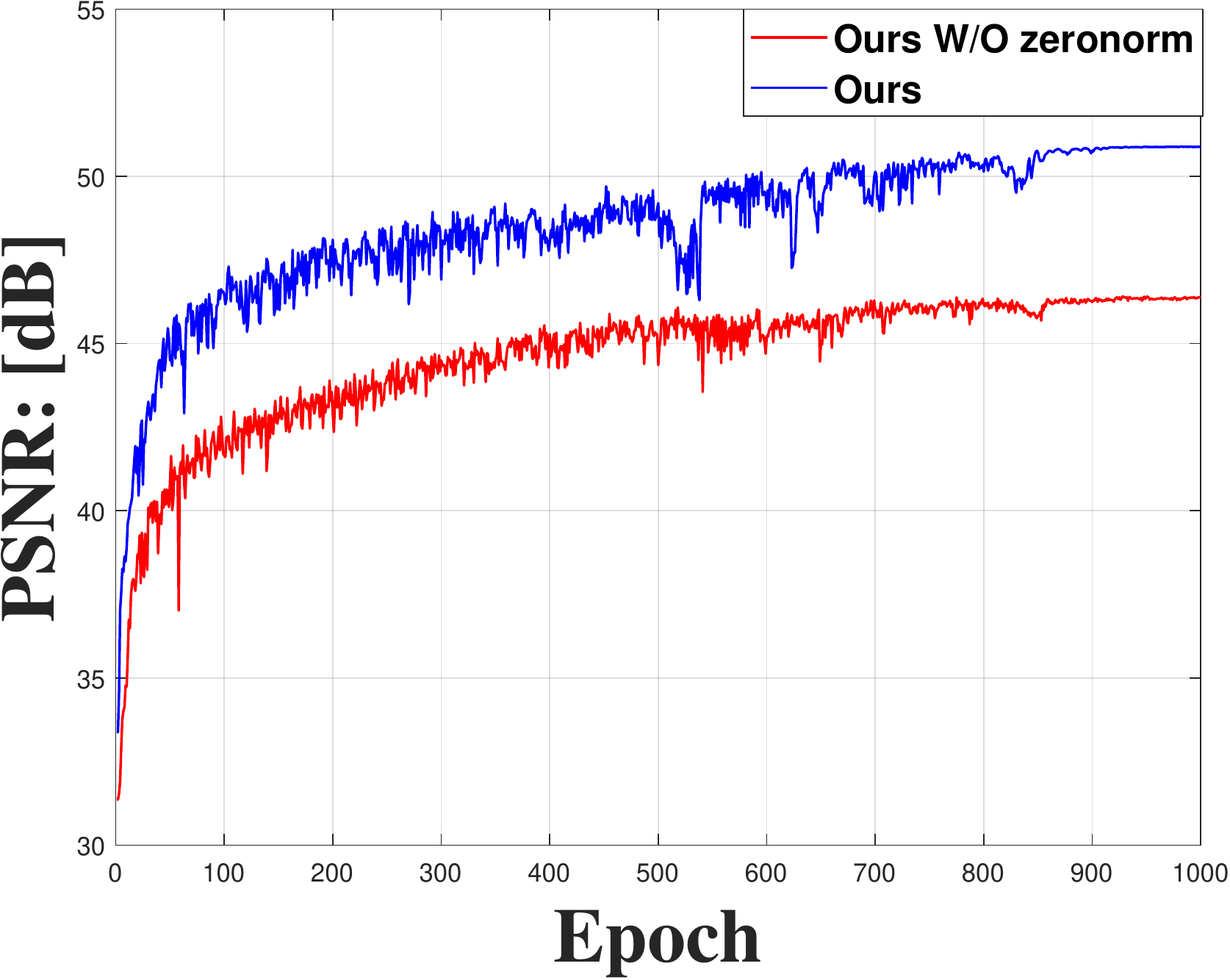}}
	\subfigure[Training Loss]{\includegraphics[width=0.45\linewidth,trim=0 0 0 0,clip]{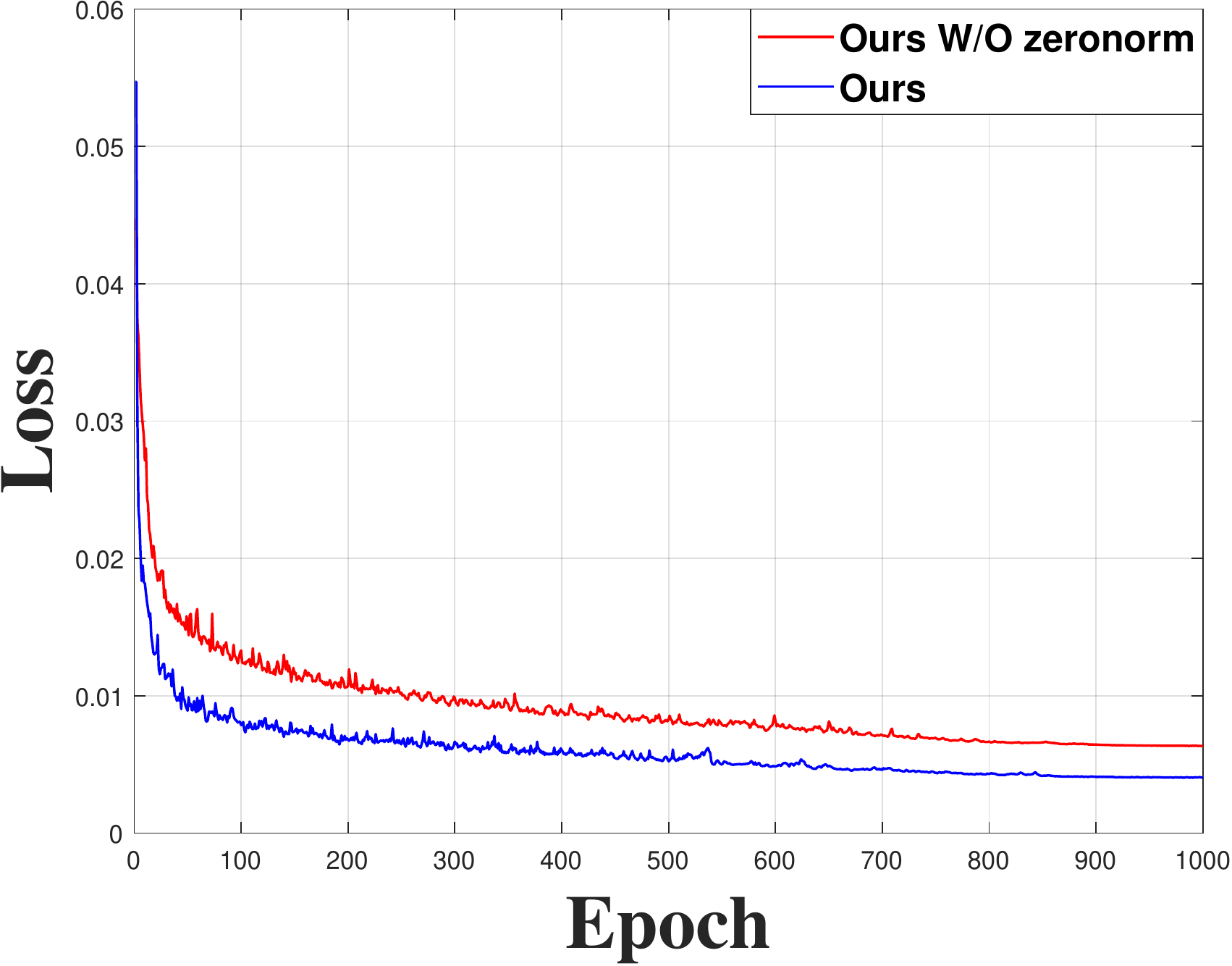}}
	\caption{The introduced advantages of our ZM-norm by illustrating the behaviors of the testing PSNR and training loss under different epochs of our PZRes-Net with or w/o ZM-norm.}
	\label{process}
\end{figure}

\subsubsection{ZM-norm and mean-value invariant up-sampling}
To predict a zero-mean residual image, ZM-norm is applied on feature maps. Also, the input LR-HSI is up-sampled with a mean-value invariant up-sampling process to avoid distortion. Here we experimentally verified the necessarily of such a combination by either removing ZM-norm during training or using a learned restriction-free transposed convolutional layer w/o the mean-value invariant property for up-sampling. From Table \ref{normalization}, we can observe
that our method achieves the best performance when both the ZM-norm and mean-value invariant up-sampling were applied. By comparing the $3^{rd}$ and $4^{th}$ rows, we can see that the ZM-norm affects the performance of our framework severely because our PZRes-Net is built upon the classic wavelet decomposition-based fusion method which focuses on extraction of the zero-mean high-frequency residual image, and  without ZM-norm, it is hard to maintain this unique property of the high-order residual image, thus leading poor performance.  Meanwhile, by comparing the $2^{nd}$ and $4^{th}$ rows, we can conclude that the mean-value invariant characteristic of the up-sampling process is necessary because without such a property, it is hard to keep the mean-values of spectral bands close to the ground-truth ones, and thus distortion is introduced. Last but not least, Fig. \ref{process} also indicates that ZM-norm can accelerate the training process.

\subsubsection{The choice of the up-sampling process}
We used the bi-linear operator to realize the  mean-value invariant up-sampling for its simplicity. We also investigated another mean-value invariant interpolation operator, i.e., the bi-cubic interpolation operator.
Experimental results in Table \ref{interpolation} indicate that  PZRes-Net with the bi-cubic and bi-linear interpolations achieve comparable performance, demonstrating the robustness of our framework. Compared with the bi-cubic interpolation, the bi-linear interpolation is more computationally efficient.
\begin{figure}
    \centering
	\subfigure[]{\includegraphics[width=0.49\linewidth]{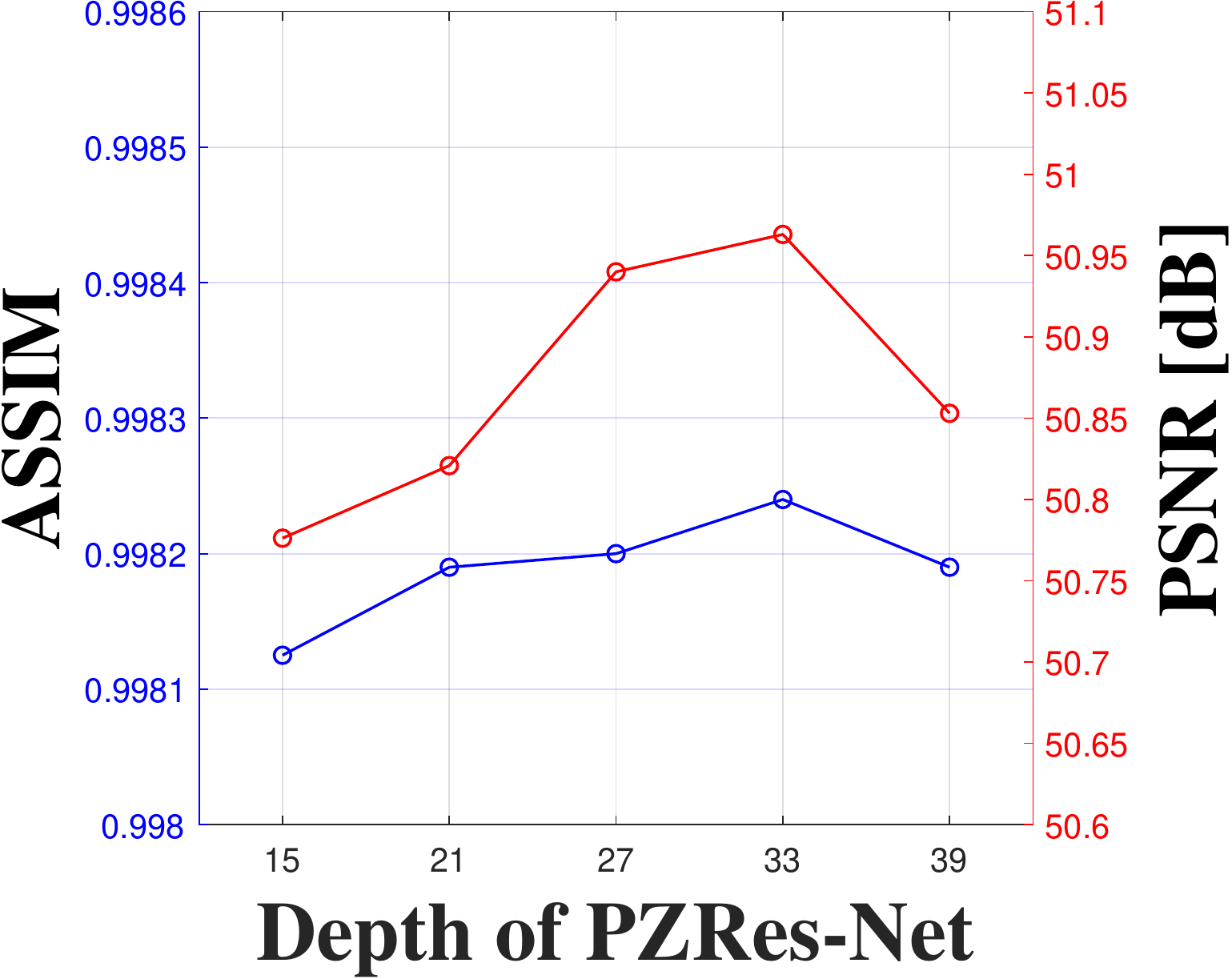}}
	\subfigure[]{\includegraphics[width=0.49\linewidth]{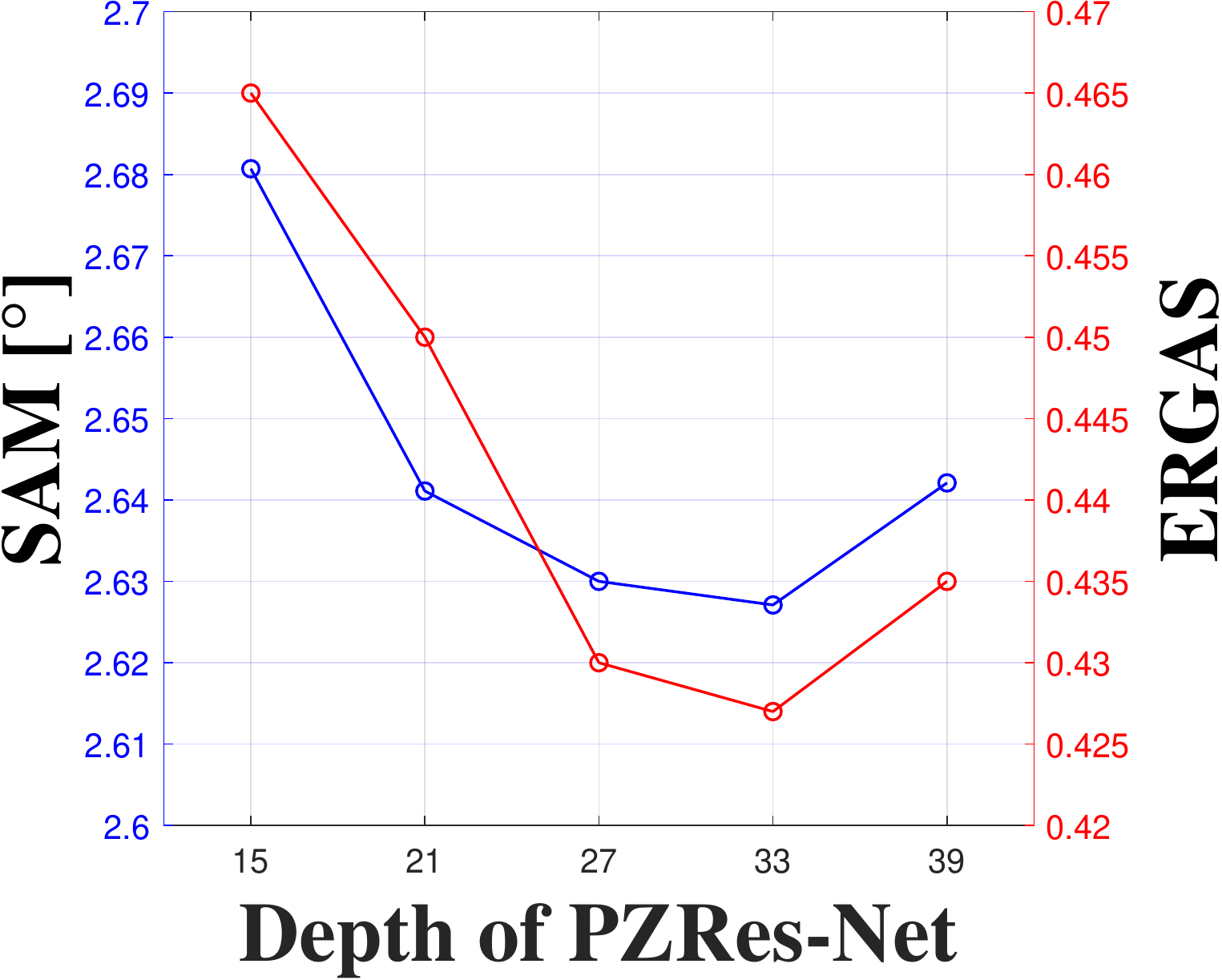}}
    \caption{Performance evaluation of the proposed method with different depths. The depth refers to the total number of 3S convolutional layers in PZRes-Net, i.e., the sum of  3S convolutional layers in all the three stages and the refinement module. Different stages have the same number of layers, and the number of 3S convolutional layers in the refinement module was fixed to 3. }
    \label{fig:depth}
\end{figure}

\subsubsection{The depth of the proposed PZRes-Net}
 We varied the total number of 3S convolutional layers in PZRes-Net from 15 to 39 with a regular interval of 6 to evaluate how the network depth affects the overall performance as well as the overfitting risk of our method.  
As shown in Fig. \ref{fig:depth}, we can see that the performance of our method in terms of four metrics gradually improves with the depth increasing. However, when the depth rises from 33 to 39, the performance decreases, indicating that overfitting of our method on the training dataset occurs. 

\section{Conclusions}
\label{sec:CON}
In this paper, we have presented a progressive zero-centric residual network (PZRes-Net), which is capable of efficiently and effectively restoring HR-HSIs from hybrid inputs, including an LR-HSI and an HR-MSI. Our PZRes-Net is mainly inspired by the classic wavelet decomposition-based image fusion method and mimics it in an adaptive learning manner. That is, our PZRes-Net mainly aims to learn a zero-centric residual image from both inputs, which contains high-frequency spatial details of the scene all spectral bands. we have proposed using ZM-norm, mean-value invariant up-sampling, spectral-spatial separable convolution with dense aggregation, and progressive spectral information embedding to achieve the objective. Extensive experimental results as well as comprehensive ablation studies on both synthetic and real benchmark datasets demonstrate that our PZRes-Net improves the current state-of-the-art performance to a new level both quantitatively and qualitatively. Moreover, our PZRes-Net is a lightweight network, which is much more computationally efficient than state-of-the-art deep learning-based methods, which validates it practicality. 

Encouraged by the impressive reconstruction quality, it raises our interests to investigate the potential of our zero-centric residual learning scheme in other high-order feature extraction tasks, e.g., reference-based image super-resolution.

\balance
\bibliographystyle{IEEEtran}
\bibliography{ref}

\begin{IEEEbiography}[{\includegraphics[width=1in,height=1.25in,clip,keepaspectratio]{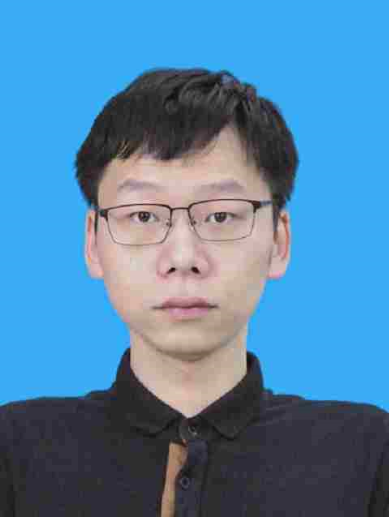}}]{Zhiyu Zhu}
received the B.E. and M.E. degrees in Mechatronic Engineering, both from Harbin Institute of Technology, in 2017 and 2019, respectively. He is currently pursuing the Ph.D. degree in computer science with the City University of Hong Kong. His research interests include hyperspectral image processing and deep learning.
\end{IEEEbiography}

\begin{IEEEbiography}[{\includegraphics[width=1in,height=1.25in,clip,keepaspectratio]{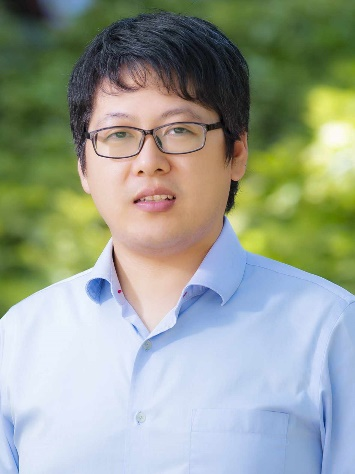}}]{Junhui Hou}(S'13-M'16-SM'20)
received the B.Eng. degree in information engineering (Talented Students Program) from the South China University of Technology, Guangzhou, China, in 2009, the M.Eng. degree in signal and information processing from Northwestern Polytechnical University, Xi'an, China, in 2012, and the Ph.D. degree in electrical and electronic engineering from the School of Electrical and Electronic Engineering, Nanyang Technological University, Singapore, in 2016. He has been an Assistant Professor with the Department of Computer Science, City University of Hong Kong, since 2017. His research interests fall into the general areas of visual computing, such as image/video/3D geometry data representation, processing and analysis, semi/un-supervised data modeling, and data compression and adaptive transmission

Dr. Hou was the recipient of several prestigious awards, including the Chinese Government Award for Outstanding Students Study Abroad from China Scholarship Council in 2015, and the Early Career Award (3/381) from the Hong Kong Research Grants Council in 2018. He is a member of Multimedia Systems \& Applications Technical Committee (MSA-TC), IEEE CAS. He is currently serving as an Associate Editor for IEEE Transactions on Circuits and Systems for Video Technology, The Visual Computer, and Signal Processing: Image Communication, and the Guest Editor for the IEEE Journal of Selected Topics in Applied Earth Observations and Remote Sensing. He also served as an Area Chair of ACM MM 2019 and 2020, IEEE ICME 2020, and WACV 2021. He is a senior member of IEEE.

\end{IEEEbiography}


\begin{IEEEbiography}[{\includegraphics[width=1in,height=1.25in,clip,keepaspectratio]{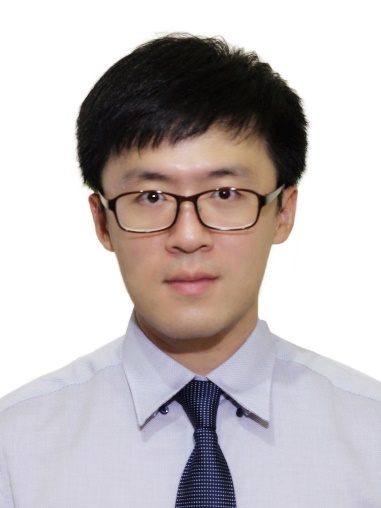}}]{Jie Chen} (S’13-M’16)
is currently an Assistant Professor at the Department of Computer Science, Hong Kong Baptist University. Before joining HKBU, he was a senior algorithm engineer at OmniVision Technologies Inc., Singapore. He worked as post-doctoral research fellow at ST Engineering-NTU Corporate Laboratory, Singapore for three years after he received Ph.D. degree in the School of Electrical \& Electronic Engineering, Nanyang Technological University, Singapore. He received B.Sc. and M. Eng. degrees both from School of Optical and Electronic Information, Huazhong University of Science and Technology, China. His research focuses on mathematical and machine learning models that solve computational photography and computer vision problems. He currently serves as an Associate Editor for the Visual Computer Journal, Springer.
\end{IEEEbiography}

\begin{IEEEbiography}[{\includegraphics[width=1in,height=1.25in,clip,keepaspectratio]{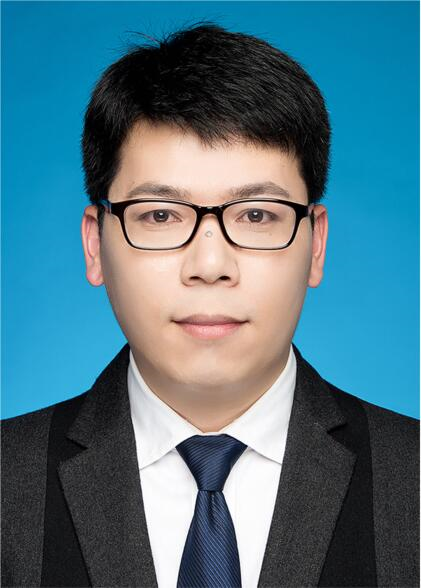}}]{Huanqiang Zeng}(S'10-M'13-SM'18)
received the B.S. and M.S. degrees from Huaqiao University, Xiamen, China and the Ph.D. degree from Nanyang Technological University, Singapore, all in electrical engineering.

He is currently a Full Professor at the School of Information Science and Engineering, Huaqiao University, Xiamen, China. Before that, he was a Postdoctoral Fellow at the Chinese University of Hong Kong, Hong Kong. His research interests include image processing, video coding, machine learning, and computer vision. He has published more than 100 papers in well-known journals and conferences, including three best poster/paper awards (in International Forum of Digital TV and Multimedia Communication 2018, and in Chinese Conference on Signal Processing 2017/2019), etc. He has been actively serving as the Associate Editor for IEEE Transactions on Image Processing, IEEE Transactions on Circuits and Systems for Video Technology, and IET Electronics Letters, the Guest Editor for Journal of Visual Communication and Image Representation, Multimedia Tools and Applications, and Journal of Ambient Intelligence and Humanized Computing. He has also been actively serving as the General Co-Chair for IEEE International Symposium on Intelligent Signal Processing and Communication Systems 2017 (ISPACS2017), the Co-Organizers of ICME2020 Workshop on 3D Point Cloud Processing, Analysis, Compression, and Communication, the Technical Program Co-Chair for Asia-Pacific Signal and Information Processing Association Annual Summit and Conference 2017 (APSIPA-ASC2017), the Area Chair for IEEE International Conference on Visual Communications and Image Processing (VCIP2015 \& VCIP2020), and the Technical Program Committee Member for multiple flagship international conferences. He is a senior member of IEEE.
\end{IEEEbiography}

\begin{IEEEbiography}[{\includegraphics[width=1in,height=1.25in,clip,keepaspectratio]{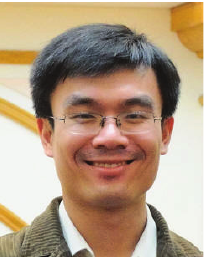}}]{Jiantao Zhou}(SM’19)
Jiantao Zhou received the B.Eng. degree from the Department of Electronic Engineering, Dalian University of Technology, in 2002, the M.Phil. degree from the Department of Radio Engineering, Southeast University, in 2005, and the Ph.D. degree from the Department of Electronic and Computer Engineering, Hong Kong University of Science and Technology, in 2009. He held various research positions with University of Illinois at Urbana-Champaign, Hong Kong University of Science and Technology, and McMaster University. He is an Associate Professor with the Department of Computer and Information Science, Faculty of Science and Technology, University of Macau, and also the Interim Head of the newly established Centre for Artificial Intelligence and Robotics. His research interests include multimedia security and forensics, multimedia signal processing, artificial intelligence and big data. He holds four granted U.S. patents and two granted Chinese patents. He has co-authored two papers that received the Best Paper Award at the IEEE Pacific-Rim Conference on Multimedia in 2007 and the Best Student Paper Award at the IEEE International Conference on Multimedia and Expo in 2016. He is an Associate Editor of the IEEE TRANSACTIONS on IMAGE PROCESSING.

J. T. Zhou is with the State Key Laboratory of Internet of Things for Smart City, and also with the Department of Computer and Information Science, University of Macau. 
\end{IEEEbiography}
\end{document}